\documentclass[useAMS,usenatbib]{mn2e}
\usepackage{graphicx, times}
\usepackage{lscape}
\IfFileExists{url.sty}{\usepackage{url}}
                      {\newcommand{\url}{\texttt}}

\providecommand{\tabularnewline}{\\}
\newcommand{\aap}{A\&A}
\newcommand{\apj}{ApJ}

\newcommand{\aj}{AJ}
\newcommand{\mnras}{MNRAS}
\newcommand{\apjs}{ApJS}

\title[Simulated observations of a young stellar cluster]{Synthetic infrared images and spectral energy distributions of a young low-mass stellar cluster}

\author[R. Kurosawa et\,al.]{Ryuichi Kurosawa\thanks{E-mail: rk@astro.ex.ac.uk}, Tim J. Harries, Matthew R. Bate and Neil H. Symington\\ School of Physics, University of Exeter, Stocker Road, Exeter EX4 4QL}

\begin{document}
\date{Dates to be inserted}

\pagerange{\pageref{firstpage}--\pageref{lastpage}} \pubyear{2004}

\maketitle

\label{firstpage}

\begin{abstract}We present three-dimensional Monte Carlo radiative
  transfer models of a very young ($<10^{5}$ years old) low mass
  ($50\,\mathrm{M_{\sun}}$) stellar cluster containing 23 stars and 27
  brown dwarfs. The models use the density and the stellar mass
  distributions from the large-scale smoothed particle hydrodynamics
  (SPH) simulation of the formation of a low-mass stellar cluster by
  Bate, Bonnell and Bromm. Using adaptive mesh refinement, the SPH
  density is mapped to the radiative transfer grid without loss of
  resolution. The temperature of the ISM and the circumstellar dust is
  computed using Lucy's Monte Carlo radiative equilibrium algorithm.
  Based on this temperature, we compute the spectral energy
  distributions of the whole cluster and the individual objects.  We
  also compute simulated far-infrared Spitzer Space Telescope
  (\emph{SST}) images (24, 70, and 160~$\mathrm{\mathrm{\mu m}}$
  bands) and construct colour-colour diagrams (near-infrared
  \emph{HKL} and \emph{SST} mid-infrared bands). The presence of
  accretion discs around the light sources influences the morphology
  of the dust temperature structure on a large scale (up to a several
  $10^{4}$~au). A considerable fraction of the interstellar dust is
  underheated compared to a model without the accretion discs because
  the radiation from the light sources is blocked/shadowed by the
  discs. The spectral energy distribution (SED) of the model cluster
  with accretion discs shows excess emission at
  $\lambda=$~3--30~$\mathrm{\mu m}$ and $\lambda>500\,\mathrm{\mu m}$,
  compared to that without accretion discs. While the former is caused
  by the warm dust present in the discs, the latter is cause by the
  presence of the underheated (shadowed) dust. Our model with
  accretion discs around each object shows a similar distribution of
  spectral index (2.2--20~$\mathrm{\mathrm{\mu m}}$) values (i.e.
  Class 0--III sources) as seen in the $\rho$~Ophiuchus cloud. We confirm
  that the best diagnostics for identifying objects with accretion
  discs are mid-infrared ($\lambda=$ 3--10~$\mathrm{\mu m}$)
  colours (e.g. \emph{SST}~IRAC bands) rather than \emph{HKL}
  colours.

\end{abstract}


\section{Introduction}

Systematic investigations of young stellar objects (YSOs) in a
star-forming cloud including comparative studies of theoretical
predictions and observations are important for understanding the
stellar/substellar formation processes. Examples of well-studied
star-forming clouds are the Orion Trapezium Cluster, NGC~2024, and the
$\rho$~Ophiuchus and Taurus-Auriga clouds. Recent observations of
young clusters have been used to determine circumstellar disc
frequencies, disc mass distributions, initial mass functions (IMFs)
and evolutional stages of the objects in the clusters. It has been
shown that \emph{JHKL} and \emph{HKL} colour-colour diagrams are particularly
effective in identifying the presence of circumstellar discs (e.g.
\citealt{kenyon:1995}; \citealt{McCaughrean:1995};
\citealt{lada:2000}; \citealt{haisch:2001}).  The spectral indices
\citep{lada:1987} or the slopes of observed spectral energy
distributions (SEDs) of YSOs from near- to mid-infrared wavelengths
are used to classify SEDs, and the classification scheme is often
related to the evolutionary stages of YSOs (e.g. \citealt*{adam:1987};
\citealt{myers:1987}).  The distribution of circumstellar disk masses
in young clusters can be measured from millimetre continuum emission
(e.g.  \citealt{eisner:2003} for NGC~2024). Determining IMFs of young
clusters requires evolutionary models (e.g. \citealt{d-antona:1997};
\citealt{baraffe:1998})
and infrared spectroscopy for spectral type identifications (e.g.
\citealt{luhman:1999} for the $\rho$~Ophiuchus cloud).

\citet*{bate:2003a} presented results from a very large
three-dimensional (3-D) smoothed particle hydrodynamics (SPH)
simulation of the collapse and fragmentation of a
50\,$\mathrm{M_{\sun}}$ turbulent molecular cloud to form a stellar
cluster. The calculation resolved circumstellar discs down to
$\sim10$~au in radius and binary stars as close as 1~au.  Although
some observational predictions, such as the IMF and binary fraction,
may be gleaned directly from a hydrodynamical simulation of stellar
cluster formation, the principal observable characteristics (optical,
near-IR, IR, and sub-millimetre images and spectra) require further
detailed radiative transfer modelling. The density distribution of
such hydrodynamical calculations is very complicated, and the
corresponding radiative transfer must be also performed in full 3-D.

There are two basic approaches to 3-D radiative transfer problems:
\,grid based methods (e.g. finite differencing, short- and long-
characteristic methods), and particle (photon) based methods, i.e.
Monte Carlo. Examples of the first kind are \citet*{stenholm:1991},
\citet{folini:2003} and \citet*{steinacker:2002}. Those of the second
kind include \citet{witt:1996}, \citet{pagani:1998}, \citet*{wolf:1999},
\citet{harries:2000} and \citet{kurosawa:2001a}. The advantages
of the second approach are, for example, the flexibility to treat
a complex density distribution and a complex scattering function.
Readers are referred to \citet{steinacker:2003} and \citet{pascucci:2003}
for more extensive discussion on the advantages and disadvantages
of these two different methods. 

The temperature of the interstellar and circumstellar dust in the
cluster must be calculated in order to determine the source function
of the dust emission. The radiative equilibrium temperature of the
dust particles can be found using the Monte Carlo method (e.g.
\citealt*{lefevre:1982}; \citealt{wolf:1999}; \citealt{lucy:1999b};
and \citealt{bjorkman:2001}).  The technique used by
\citet{lucy:1999b} takes into account the fractional photon
absorptions between two events of a `photon packet'; hence, it works
well even in the limit of low opacity. \citet{bjorkman:2001} used the
immediate re-emission technique, in which radiative equilibrium is
forced at each interaction with the dust. A photon is re-emitted
immediately after an absorption event using a product of the dust
opacity and the difference between the Planck function with a current
temperature and that with a new temperature corresponding to the
radiative equilibrium of the dust that absorbed the photon. 
Unlike the method of 
\citet{lucy:1999b}, this does not require a temperature iteration if
the opacity is independent of temperature. Alternatively,
\citet*{niccolini:2003} used the ray-splitting method showing its
effectiveness for both low and high optical depth media.

Here we aim to simultaneously resolve the dust on parsec and
sub-stellar-radius spatial scales, whilst including multiple radiation
sources. To overcome the resolution problem, we have implemented an
adaptive mesh refinement (AMR) scheme in the grid production process
of the TORUS radiative transfer \citep{harries:2000}. See also
\citet{wolf:1999}, \citet{kurosawa:2001a} and \citet{steinacker:2002}
for a similar gridding scheme in a radiative transfer problem. The
method described by \citet{lucy:1999b} is used to compute dust
temperatures in our models.

The objectives of this paper are: 1.\,to compute observable quantities
by solving the radiative transfer problem using a complex density
distribution from the SPH calculation by \citet{bate:2003a}; 2.\,to
analyse the predicted observational properties of the cluster
generated in the simulation of \citet{bate:2003a} at a distance of
140~pc which corresponds to the distance of nearby star-forming
regions such as Taurus-Auriga and the $\rho\,$Ophiuchus cloud (e.g.
\citealt*{bertout:1999}).

In Section~2, we describe the details of our models. The results of
the model calculations are given in Section~3. The conclusions are
summarised in Section~4.

\begin{figure}

\begin{center}

\includegraphics[%
  clip,
  scale=0.45]{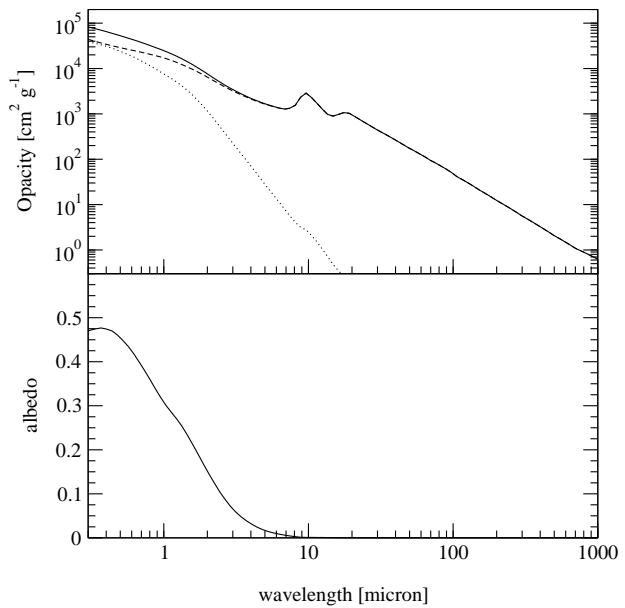}

\end{center}

\caption{Top: The scattering (dotted line), the absorption (dashed
line), and the total (solid line) opacities of the grain model,
described in Section~\ref{sub:Dust-Model}, are shown as functions of
wavelength. Bottom: The corresponding albedo of the grains is shown as
a function of wavelength.} 

\label{fig:albedo}

\end{figure}

\section{Methods\label{sec:Models}}

We calculate the main observable quantities of the low mass (50~$\mathrm{M_{\sun}}$)
cluster formation simulation by \citet{bate:2003a}. There are four
basic steps in this process: 

\begin{enumerate}
\item Create the AMR grid from the SPH particle positions and the density
at each particle. The density is assigned to the grid cells during
the grid construction. 
\item Calculate the luminosity and the effective temperature of each light
source (in the SPH calculation) based on its mass using an evolutional
model of young stars (\citealt{d-antona:1998}; \citealt{censori:1998}). 
\item Compute the temperature of the dust in the cluster using the Monte
Carlo radiative equilibrium model of \citet{lucy:1999b}. 
\item Compute the SEDs and the images using the Monte Carlo radiative transfer
code, TORUS. 
\end{enumerate}

\subsection{The SPH model }

Bate et al.~(\citeyear{bate:2002a}, \citeyear{bate:2002b},
\citeyear{bate:2003a}) presented a numerical SPH simulation of star
cluster formation resolving the fragmentation process down to the
opacity limit ($\sim$0.005\,M$_\odot$). The initial conditions consist
of a large-scale, turbulent molecular cloud with a mass of
50~$\mathrm{M_{\sun}}$ and a diameter of 0.375~pc (77\,400~au).  The
initial temperature of the cloud was 10~K, and its corresponding mean
thermal Jeans mass was 1~$\mathrm{M_{\sun}}$ (i.e. the cloud contained
50 thermal Jeans masses). The free-fall time of the cloud was
$t_{\mathrm{ff}}=6.0\times10^{12}\,\mathrm{{s}}=1.90\times10^{5}\,\mathrm{{yr}}$.
Similar to the method used by \citet*{ostriker:2001}, they imposed an
initial supersonic turbulent velocity field on the cloud by generating
a divergence-free random Gaussian velocity field with a power spectrum
$P\left(k\right)=k^{-4}$ where $k$ is the wave number. This was chosen
to reproduce the observed Larson scaling relations \citep{larson:1981}
for molecular clouds. The total number of particles used in the
simulations was $3.5\times10^{6}$, making it one of the largest SPH
calculations ever performed. Approximately 95\,000 CPU hours on the
SGI Origin 3800 of the United Kingdom Astrophysical Fluids Facility
(UKAFF) were spent on the calculation.

We use the output from the final time step of the SPH calculation as
the input for the radiative transfer code. The time of the data dump
is 0.27~Myr, at which the cluster contains 50 point masses (stars and
brown dwarfs). Each SPH data point contains the position ($x$, $y$,
$z$), the velocity components ($V_{x}$, $V_{y}$, $V_{z}$) and the
density ($\rho$). The velocity information is not used in our
calculation since we are only interested in continuum emission, but
these data may be used in future calculations of molecular line
emission.  The total masses contained in the stars and the molecular gas
are $6\,\mathrm{M_{\sun}}$ and $44\,\mathrm{M_{\sun}}$ respectively.

\subsection{Source catalogue}

The SPH data provides the masses of the stellar objects (23 stars and
27 brown dwarfs), ranging from 0.005 to 0.731 $\mathrm{M_{\sun}}$ (see
Table~\ref{tab:catalog}). Since the Monte Carlo radiative transfer
code requires the luminosity ($L$), the effective temperature
($T_{\mathrm{{eff}}}$) and the radius ($R_*$) of each star, we
computed them indirectly from evolutionary models. For a given age and
mass of each star and brown dwarf, the luminosity and the temperature
are interpolated from the 1998 updated version of data by
\citet{d-antona:1998} and \citet{censori:1998} available on their
website (\url{http://www.mporzio.astro.it/~dantona/prems.html}).
Although the actual age of the stars and brown dwarfs in the SPH data
of \citet{bate:2003a} ranges from $\sim$2\,000 to $\sim$70\,000 years
old, we make the pragmatic assumption that all the objects are
0.25~Myr old because we believe that at young ages the stellar radii
are overestimated (c.f., \citealt{baraffe:2002}), leading to
unrealistically high luminosities. Adopting models with ages of
0.25\,Myr provides more plausible radii (and therefore luminosities),
while the change in temperature over this timespan is modest
(\citealt{baraffe:2002}).

The radii are estimated from $L=4\pi R^{2}\sigma
T_{\mathrm{{eff}}}^{4}$ where $\sigma$ is the Stefan-Boltzmann
constant.  The results are summarised in Table~\ref{tab:catalog}.  The
combined luminosity of all 50 sources is $16.8\,\mathrm{L_{\sun}}$.

We take the spectral energy distribution of the individual stars to be
blackbodies. Although we recognise that the spectra of the objects
(particularly at the lowest masses) may be significantly structured
due to molecular opacity, the vast majority of the stellar flux is
reprocessed by the circumstellar dust and the precise form of the
input spectrum is unimportant.

\begin{table*}

\vbox to220mm{\vfil Landscape table to go here. (Table 1 attached to the end of paper.)

\caption{}

\vfil}

\end{table*}

\begin{figure*} 

\begin{center}

\begin{tabular}{rrr}
\includegraphics[%
  scale=0.7]{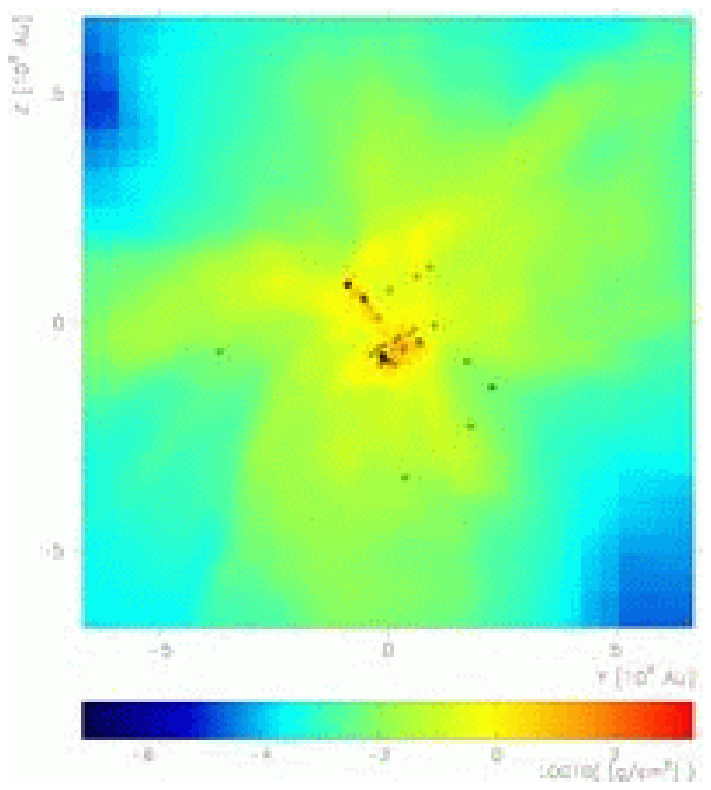}&
\includegraphics[%
  scale=0.7]{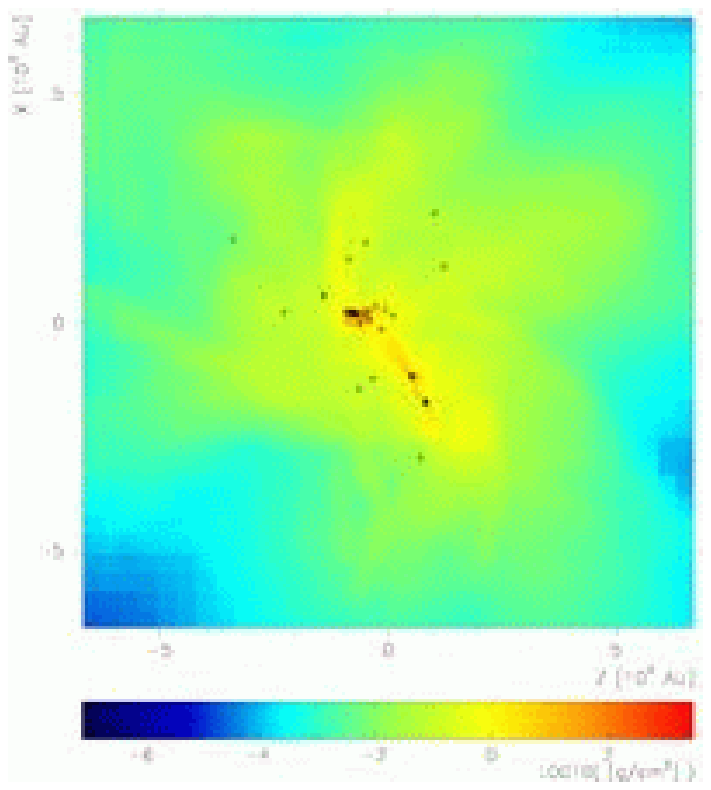}&
\includegraphics[%
  scale=0.7]{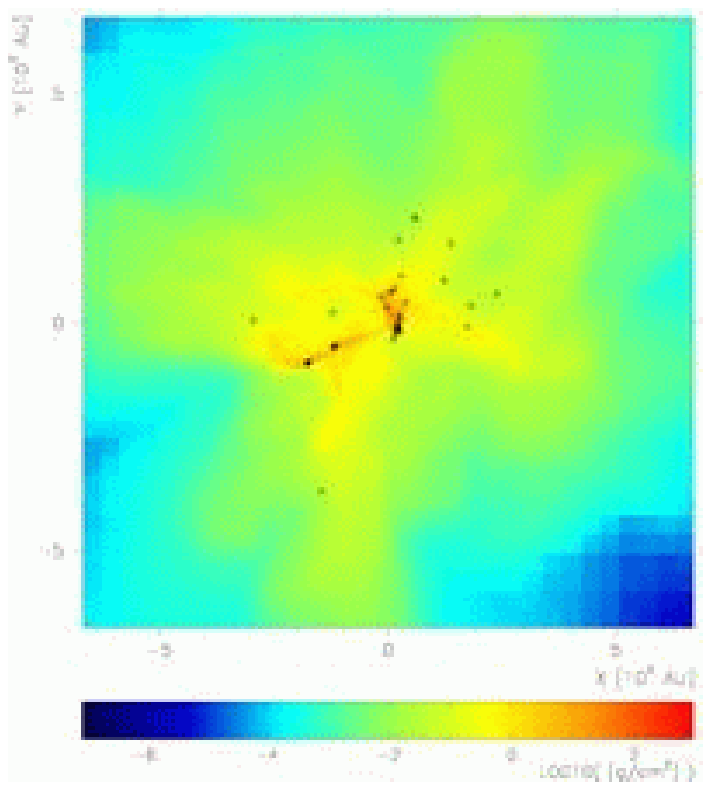}\tabularnewline
\includegraphics[%
  scale=0.7]{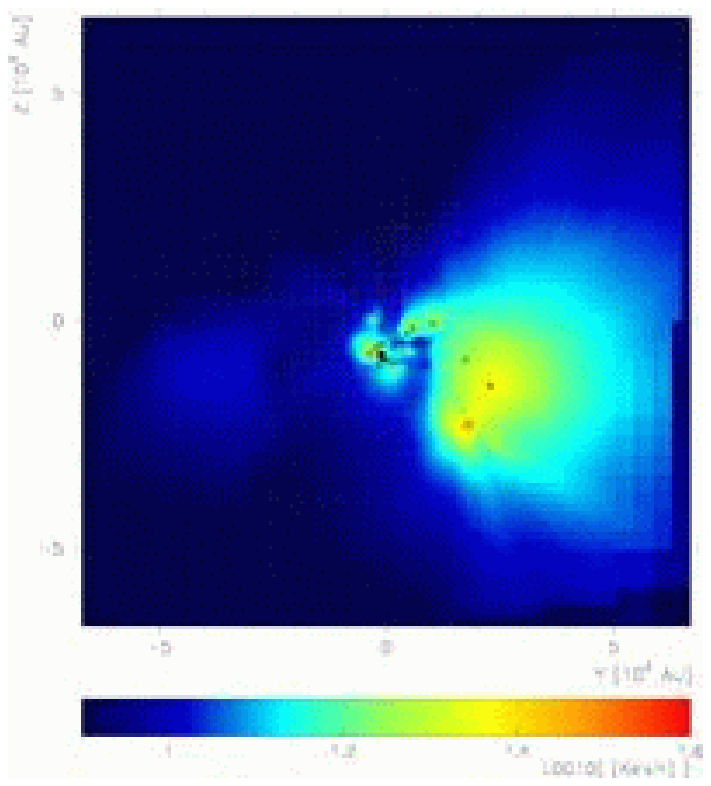}&
\includegraphics[%
  scale=0.7]{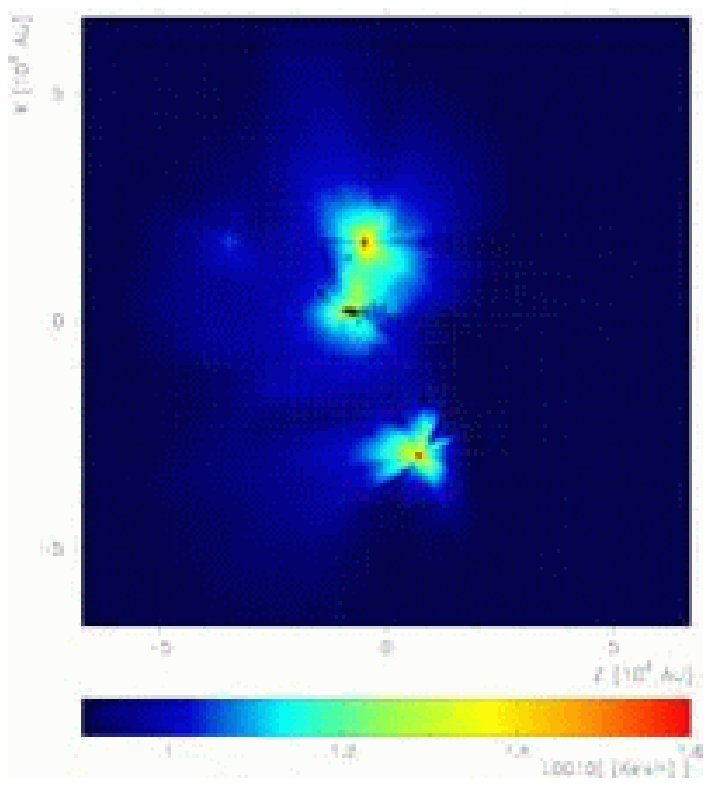}&
\includegraphics[%
  scale=0.7]{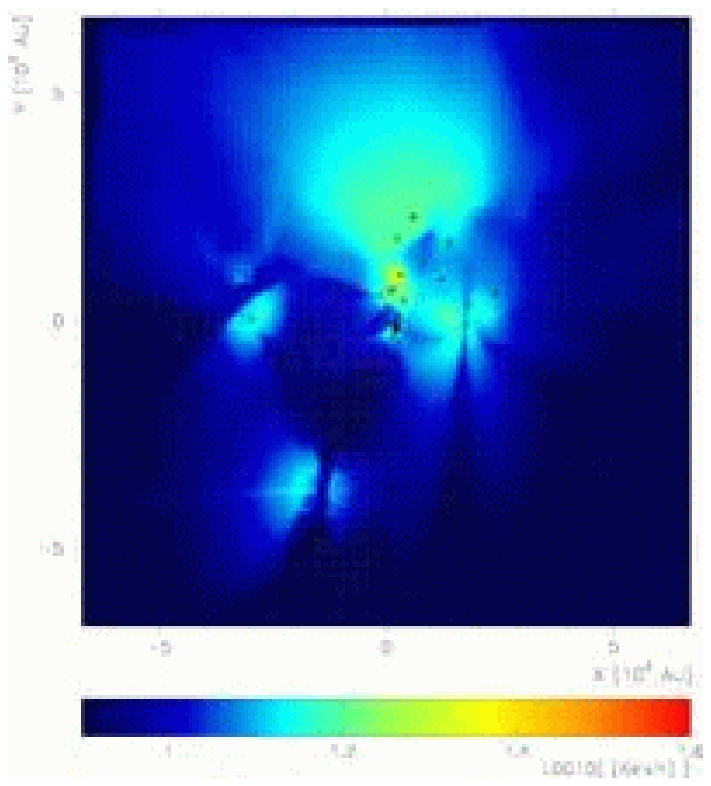}\tabularnewline
\includegraphics[%
  scale=0.625]{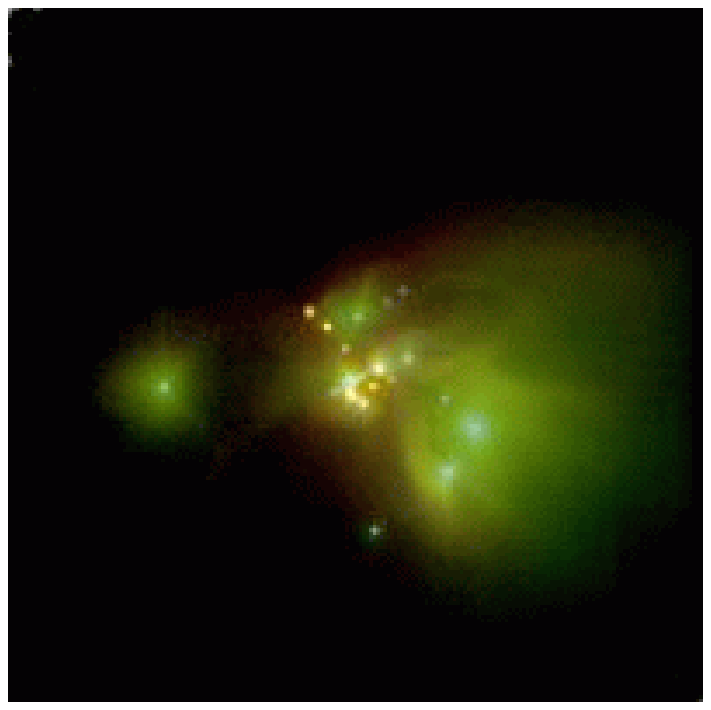}&
\includegraphics[%
  scale=0.625]{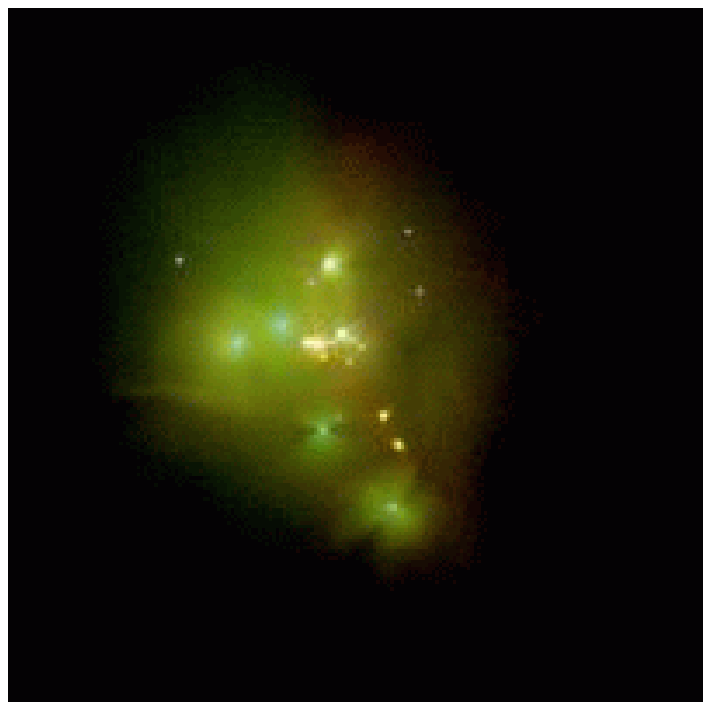}&
\includegraphics[%
  scale=0.625]{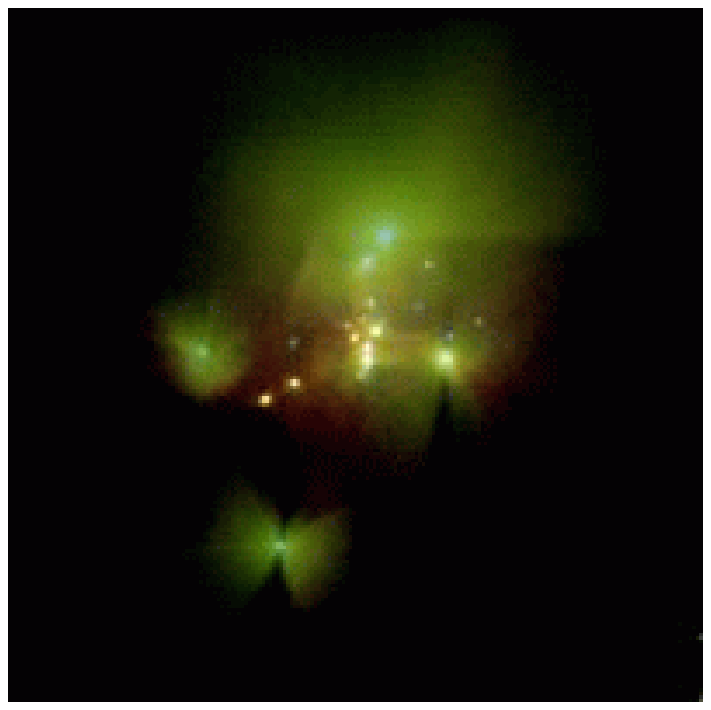}\tabularnewline
\end{tabular}

\end{center}

\caption{Top: the log of dust column density (in $\mathrm{g~cm^{-2}}$)
along $x$, $y$ and $z$ axes from left to right respectively (both Models
I and II).  Middle: The temperature on $x=0$, $y=0$, and $z=0$ planes
from left to right (Model II).  Bottom: The composite images of \em
SST \em MIPS bands -- 24~$\mathrm{\mu m}$ (blue) 70~$\mathrm{\mu m}$
(green) and 160~$\mathrm{\mu m}$ (red) --  along the x, y and z axes
from left to right respectively (Model II). The black circles in the
column density and temperature plots indicate the locations of the
stars and brown dwarfs projected on each plane. }

\label{fig:ColumnDensity}

\end{figure*}

\begin{figure*}

\begin{center}

\begin{tabular}{cc}
\includegraphics[%
  scale=1.1]{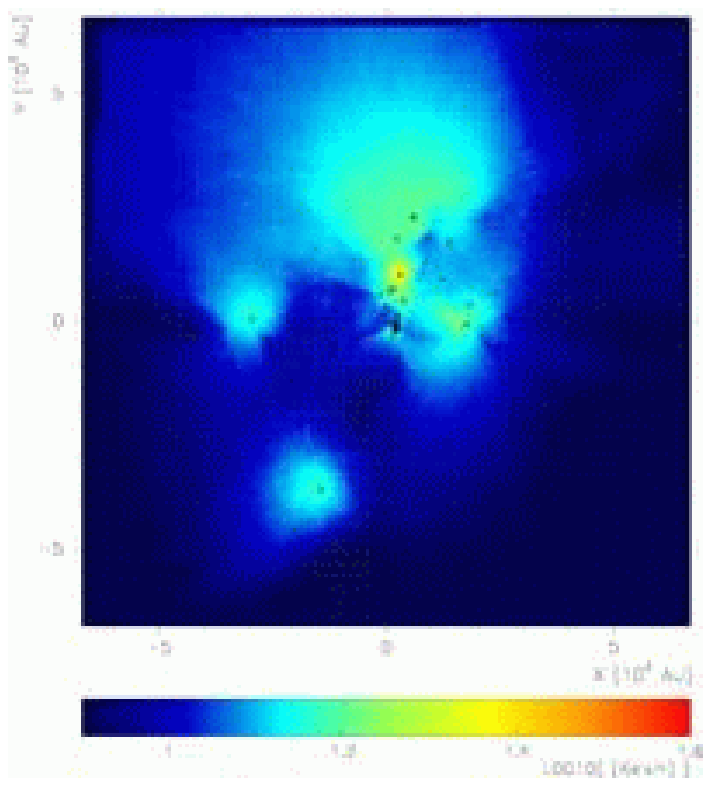}&
\includegraphics[%
  scale=1.1]{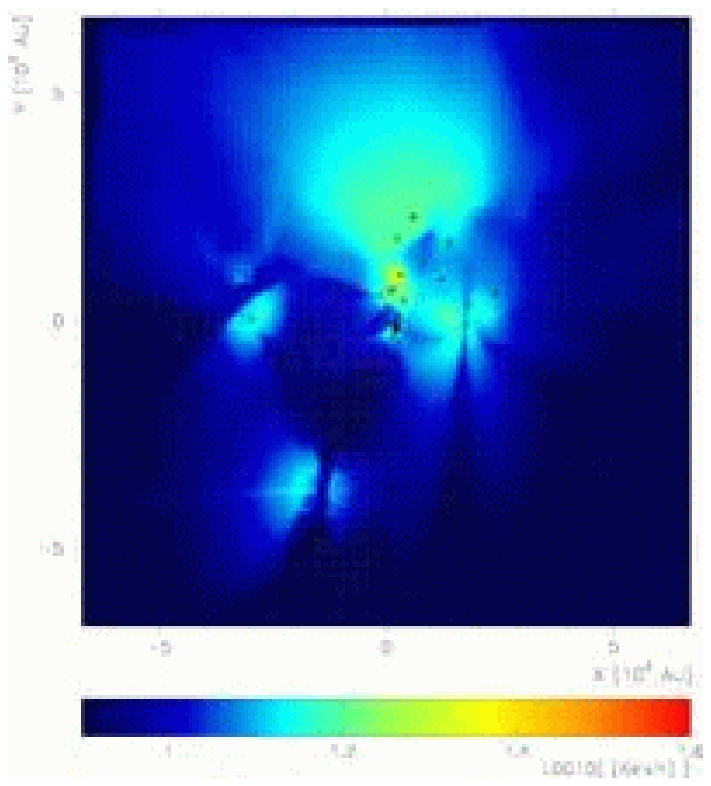}\tabularnewline
\end{tabular}

\end{center}

\caption{Log-scale temperature maps (in kelvins) from Model I (left)
  and Model II (right) on the $z=0$ plane. The circles indicate the
  positions of stars and brown dwarfs projected on the plane. The
  temperature of the dust shadowed by the accretion discs (left) is
  significantly lower than that of Model I. Although the same colour
  scale is used for both maps, the actual maximum dust temperatures
  for Model I and Model II are around 300~K and 1600~K respectively.}

\label{fig:temperature-map}

\vspace{0.5cm}

\end{figure*}

\subsection{The grid construction (AMR) \label{sub:The-Grid-Construction}}

The algorithm used to construct the grid in this paper is very similar
to that used in \citet{kurosawa:2001a}. More detailed discussion of
the AMR grid construction and the data structure used in our model are
given in \citet{symington:2004}. Starting from a large cubic cell
(with size $d$) which contains all the SPH particles, we first compute
the density of the cell by averaging the density values assigned to
the particles. Then, the density is multiplied by the volume of the
cell ($d^{3}$) to find the total mass ($M$) in the cell. If the mass
is larger than a threshold mass ($M_{\mathrm{th}}$), which is a user
defined parameter, this cell is split into 8 subcells with size $d/2$.
If the mass of the cell is less than the threshold, it will not be
subdivided. Recursively, the same procedure is applied to all the
subcells until all the cells contain a mass less than the threshold
($M<M_{\mathrm{th}}$). Our largest models split the grid on 27 levels,
i.e. they incorporate a dynamical range of $2^{27}\approx10^{8}$.
The average optical depth across a cell is about $1.5$ at $2\,\mu \mathrm{m}$
which corresponds to the peak of a blackbody radiation
curve ($B_{\nu}$) with the mean temperature ($T=2900\,
\mathrm{K}$) of the stars and brown dwarfs in our models.

\subsection{Temperature calculation \label{sub:Temperature-Calculation}}

The dust temperature is computed using the method described by \citet{lucy:1999b}.
The material is assumed to be in local thermodynamic equilibrium (LTE)
and in radiative equilibrium. The former indicates that the source
function ($S_{\lambda}$) is described solely by the Planck function,
$B_{\lambda}$, for all wavelengths $\lambda$: 

\begin{equation}
S_{\lambda}=B_{\lambda}\left(T\right)\label{eq:source-planck}
\end{equation}
where $T$ is the temperature of the dust. The latter indicates that
the total energy absorbed by a volume element per unit time is exactly
same as the total amount of energy emitted by the volume in unit time:

\begin{equation}
\int_{0}^{\infty}\kappa_{\lambda}B_{\lambda}d\lambda=\int_{0}^{\infty}\kappa_{\lambda}J_{\lambda}d\lambda\,\label{eq:rad-equi}\end{equation}
 where $\kappa_{\lambda}$ and $J_{\lambda}$ are the thermal absorption
coefficient and the specific mean intensity respectively. The expression
can be derived by integrating the radiative transfer equation, $dI_{\lambda}/dl=\rho\kappa_{\lambda}\left(S_{\lambda}-I_{\lambda}\right)$,
over all wavelengths  and solid angles ($\Omega$), using
the LTE condition (equation~\ref{eq:source-planck}) and flux
conservation (c.f., \citealt{chandrasekhar:1960}). By rewriting equation~\ref{eq:rad-equi}
using the Planck mean ($\bar{\kappa}_{\mathrm{p}}$) and the absorption
mean ($\bar{\kappa}_{J}$), it becomes:\begin{equation}
J=\frac{\bar{\kappa}_{\mathrm{p}}}{\bar{\kappa}_{J}}\, B\left(T\right)\label{eq:J-B}\end{equation}
 where $\bar{\kappa}_{\mathrm{p}}$ and  $\bar{\kappa}_{J}$ 
are defined as:\begin{eqnarray}
\bar{\kappa}_{\mathrm{p}} & \equiv & \frac{\int_{0}^{\infty}\kappa_{\lambda}B\left(T\right)_{\lambda}d\lambda}{\int_{0}^{\infty}B_{\lambda}\left(T\right)d\lambda}=\frac{\pi\int_{0}^{\infty}\kappa_{\lambda}B_{\lambda}\left(T\right)d\lambda}{\sigma T^{4}}\label{eq:mean-planck}\\
\bar{\kappa}_{J} & \equiv & \frac{\int_{0}^{\infty}\kappa_{\lambda}J_{\lambda}d\lambda}{\int_{0}^{\infty}J_{\lambda}d\lambda}=\frac{\int_{0}^{\infty}\kappa_{\lambda}J_{\lambda}d\lambda}{J}\,,\label{eq:mean-absorption}\end{eqnarray}
and the (wavelength integrated) mean intensity $J$ is \begin{equation}
J=\int_{0}^{\infty}J_{\lambda}d\lambda\,.\label{eq:mean-intensity}\end{equation}
 Using the $\Lambda$ operator (e.g. \citealt{mihalas:1978}),
the formal solution of the radiative transfer equation can be written
as: \begin{equation}
J=\Lambda\left[S\right]=\Lambda\left[B\left(T\right)\right]\,.\label{eq:formal-solution}\end{equation}
 This problem is often solved by iteration starting from some initial
temperature structure. As noted by \citet{lucy:1999b}, equation~\ref{eq:J-B}
is indeed in the form of the formal solution; hence, we can use 
an iterative scheme to obtain the temperature. Note that using
$B\left(T\right)=\sigma T^{4}/\pi$ in equation~\ref{eq:J-B}, we
have: \begin{equation}
T=\left(\frac{\pi\bar{\kappa}_{J}}{\sigma\bar{\kappa}_{\mathrm{p}}}\, J\right)^{1/4}\,.\label{eq:T-solution}\end{equation}
In addition, \citet{lucy:1999b} explicitly used flux conservation
as a constraint when computing $\bar{\kappa}_{J}J$ and $\bar{\kappa}_{\mathrm{p}}$
in each iteration step. 

Our basic iteration scheme is summarised as follows:

\begin{enumerate}
\item Using the current temperature ($T_{i}$) from the $i$-th iteration
step, estimate the values of $\bar{\kappa}_{J}J$ and $\bar{\kappa}_{\mathrm{p}}$
for each cell \emph{}using the Monte Carlo method \emph{}\citep{lucy:1999b}.
\item Evaluate the new temperature ($T_{i+1}$) using equation~\ref{eq:T-solution}.
\item Check for convergence. If the model has not converged, go back to
(i).
\end{enumerate}
The model is considered to have converged when 

\begin{equation}
\delta=\frac{\Delta\bar{T}_{i+1}-\Delta\bar{T}_{i}}{\frac{1}{2}\left(\Delta\bar{T}_{i+1}+\Delta\bar{T}_{i}\right)}<0.05\label{eq:T-convergence}\end{equation}
 where $\Delta\bar{T}_{i}=\bar{T}_{i}-\bar{T}_{i-1}$, and $\bar{T}_{i}$
is the mean temperature of all cells in the $i$-th iteration. 

\citet{harries:2004} presented test calculations for the dust temperature
using our Monte Carlo model for a spherical shell (1-D) and a disc
(2-D), and showed the results agree with those of well-established
existing codes (e.g. \citealt{ivezic:1999}; \citealt{pascucci:2003}).

\subsection{Dust model \label{sub:Dust-Model}}

To calculate the dust scattering and absorption cross-sections as
functions of wavelength, we have used the optical constants of \citet{draine:1984}
for amorphous carbon grains and \citet{hanner:1988} for silicate
grains. For simplicity, the model uses the standard interstellar medium
(ISM) power-law size distribution function (e.g, \citealt*{mathis:1977}): 

\begin{equation}
n\left(a\right)\propto a^{-q}\label{eq:grain-dist-function}\end{equation}
 for $a_{\mathrm{\mathrm{min}}}<a<a_{\mathrm{max}}$ with $q=3.5$,
$a_{\mathrm{\mathrm{min}}}=5\times10^{-3}\,\mathrm{\mathrm{\mu m}}$
and $a_{\mathrm{\mathrm{min}}}=0.25\,\mathrm{\mathrm{\mathrm{\mu m}}}$.
The relative number of each grain is assumed to be that of solar abundance,
C/H$\sim3.5\times10^{-4}$ \citep{anders:1989} and Si/H$\sim3.6\times10^{-5}$
\citep{grevesse:1993} which are similar to values found in the ISM
model of \citet{mathis:1977} and \citet*{kim:1994}. Similar abundances
were used in the circumstellar disc models of \citet{cotera:2001}
and \citet{wood:2002}. Figure~\ref{fig:albedo} shows the resulting
opacities and albedo as functions of wavelength. 

Assuming spherical dust particles, the Mie-scattering phase matrix
is pre-tabulated and used in our calculations, instead of using an
approximate scattering phase function (e.g. \citealt{henyey:1941}).
We assume the same dust size distribution in the ISM and the circumstellar
discs, i.e. the circumstellar discs do not contain larger grains.
If larger grain sizes are chosen for the circumstellar dust, the
wavelength dependency of the dust opacity (Figure~\ref{fig:albedo})
will be shallower than that of the ISM dust \citep{wood:2002}. 
As a result, this may change the slope of model SEDs in sub-millimetre
wavelength range (c.f. \citealt{beckwith:1990}; \citealt{beckwith:1991}).

\subsection{Images and SED calculation}

Once the temperature structure has converged, the source function is
known everywhere in the model space (assuming LTE). Using this
information, we simulate the observable quantities, namely the SEDs and
the images.  The Monte Carlo radiative transfer method is used again
to propagate the photons, and project them on to an observer's plane
on the sky.  The basic method used here is presented in e.g.
\cite{hillier:1991} and \citet{harries:2000}.

The following sets of filters are used in our calculations: standard
\emph{J}, \emph{H}, \emph{K} and \emph{L}; \emph{SST} Infrared Array
Camera (IRAC) at 3.6, 4.5, 5.8 and 8.0~$\mathrm{\mu m}$; the \emph{SST}
Multiband Imaging Photometer (MIPS) at 24, 70, and 160~$\mathrm{\mu m}$.

\subsection{Accretion disc model\label{sub:Accretion-disc-model}}

The SPH simulations of \citet{bate:2003a} do not resolve scale sizes
less than $\sim10$~au; hence, accretion discs very close to stars and
brown dwarfs are not included. These missing accretion discs are
potentially very important in the calculations of the SED and the
images. Although large accretion discs exist around some objects in
the SPH calculation, they do so only at distances $>10\,\mathrm{au}$
from the central object. To investigate the effect of the warm dust
very close to the objects, we insert discs using the density described
by the steady $\alpha$-disc `standard model' (\citealt{shakura:1973};
\citealt*{frank:2002}).

\begin{equation}
\rho_{\mathrm{d}}\left(r,z\right)=\Sigma\left(r\right)\,\frac{1}{\sqrt{2\pi}h\left(r\right)}e^{-\left(\frac{z}{2h\left(r\right)}\right)^{2}}\label{eq:disc-density-function}\end{equation}
 where $r$ is the cylindrical radius expressed in units of the disc
 radius $R_{\rm d}$, and  $h$ and  $z$ are the scale height and 
the distance from the disc plane, respectively.  $\Sigma$ is the surface density at the mid-plane. 
The mid-plane surface density and the scale height are
given as: 
\begin{equation}
  \Sigma\left(r\right)=\frac{5M_{\mathrm{d}}}{8\pi R_{\mathrm{d}}^{2}}\,
  r^{-3/4}  \label{eq:density-midplane}
\end{equation}
 where $M_{\mathrm{d}}$ is the disc mass.
\begin{equation}
  h\left(r\right)=0.05\, R_{\mathrm{d}}\,
  r^{9/8}\,.\label{eq:scale-height}
\end{equation}
Note that the mid-plane surface density, predicted by the irradiated disk model
 of \citet{d-alessio:1998}, has a slightly steeper radial dependency
 ($\Sigma\propto r^{-1}$). 

The inner radius of the disc is set to $R_{\mathrm{i}}=6\, R_{*}$ 
which is the assumed dust destruction radius of our model (c.f., \citealt{wood:2002}).
The disc mass, $M_{\mathrm{d}}$, is assumed to be 1/100
of the central object's mass, and the disc radius ($R_{\mathrm{d}}$) to be
10~au unless it has a binary companion. When an object is in a binary
system, the disc radius is assigned to be 1/3 of the binary separation
\citep[e.g.][]{artymowicz:1994}. Finally, the orientations of the
discs are assigned from the spin angular momentum of the objects in
the SPH simulation which keeps track of the angular momentum of the
gas accreted by the objects.

\section{Results\label{sec:Results}}

We present results from two different models: I.\,the density structure
is exactly the same as in the SPH simulation (i.e. there is no
circumstellar material within 10\,au of any star) and II.\,the density
structure is from the SPH simulation with the addition of accretion
discs as described in Section~\ref{sub:Accretion-disc-model}. 
The total number of cells used in Models I and II are 2,539,440 and 25,223,192 respectively.
The corresponding cell subdivision levels are 16 and 27 for Models I
and II, providing smallest cell sizes of 1~au and $\sim0.1\,\mathrm{R_{\sun}}$
respectively. Table~\ref{tab:model-parameters} summarises the basic
model parameters. 

\begin{table}

\caption{Basic Model Parameters}

\label{tab:model-parameters}

\begin{tabular}{lll}
\hline 
&
Model I&
Model II\tabularnewline
\hline 
Size of the largest cell&
$2\times10^{18}\,\mathrm{cm}$&
$2\times10^{18}\,\mathrm{cm}$\tabularnewline
Size of the smallest cell&
1~au&
$\sim0.1\,\mathrm{R_{\sun}}$\tabularnewline
Discs within 10 au?&
No&
Yes\tabularnewline
Number of grid cells&
2,539,440&
25,223,192\tabularnewline
AMR subdivision levels&
16&
27\tabularnewline
\hline
\end{tabular}

\vspace{0.3cm}

\end{table}

\subsection{Density and temperature maps\label{sub:Density-and-temperature}}

We assume a gas-to-dust mass ratio of 100 (c.f.,
\citealt{liseau:1995}), and assign the dust density of each cell
accordingly.  The resulting dust column-density maps along $x$, $y$
and $z$ axes for Model~II are given in the top row of
Figure~\ref{fig:ColumnDensity}. On this scale, the column density maps
for Model~I look exactly the same except that the value of the maximum
column density is lower.

The results from the radiative equilibrium temperature calculations
for Models I and II are shown in Figure~\ref{fig:temperature-map}.
Approximately 500 CPU hours were spent on the SGI Origin 3800 of the
UKAFF for computing temperatures for Model I, and 2400 CPU hours for
Model~II. Although the same colour scale is used for both models in
Figure~\ref{fig:temperature-map}, the maximum dust temperatures found in
the whole domain of the models are about 300~K and 1600~K for Models I
and II, respectively. In both models, the average temperature of the
ISM dust is around 20~K. The temperature maps for Model II on $x=$0,
$y=0$, and $z=0$ planes are also given in the middle row of
Figure~\ref{fig:ColumnDensity}. The angle averaged optical (at
$\lambda=5500$~\AA) using 600 random directions from the centre of the
cluster to the outer boundary of the models is~93 ($A_V=102$).

It is clear from the temperature maps of Model~II that the presence
of the circumstellar discs influences the morphology of the temperature
structure. Discs affect the temperature of the dust even on a large
scale (up to a several $10^{4}$~au). In Model~II, the dust shadowed
by the accretion disc has a lower temperature than that of Model I.
The radiation field near the stars becomes anisotropic/bipolar
when circumstellar discs are present. 

\begin{figure*}

\begin{center}

\begin{tabular}{cc}
\includegraphics[%
  scale=1.08]{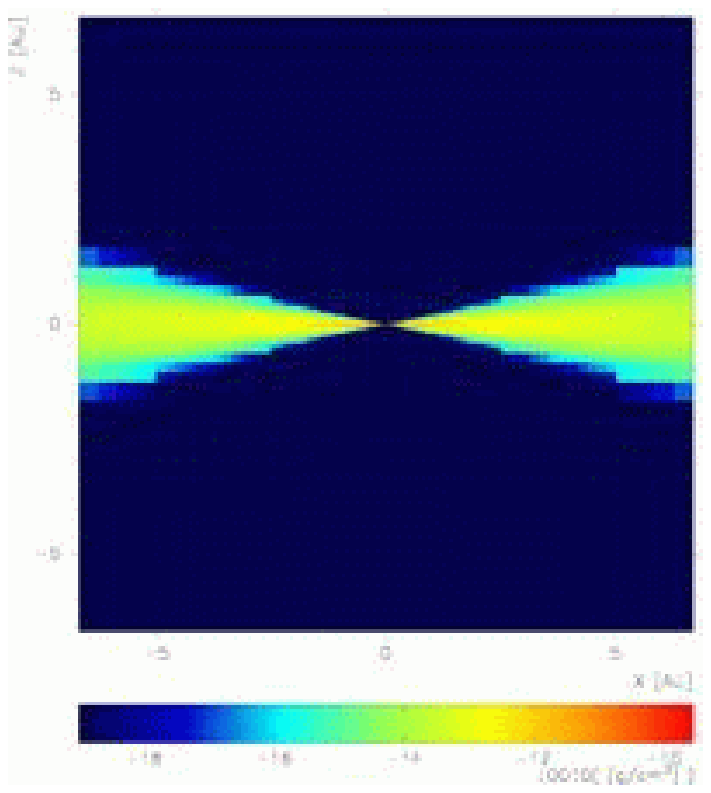}&
\includegraphics[%
  scale=1.08]{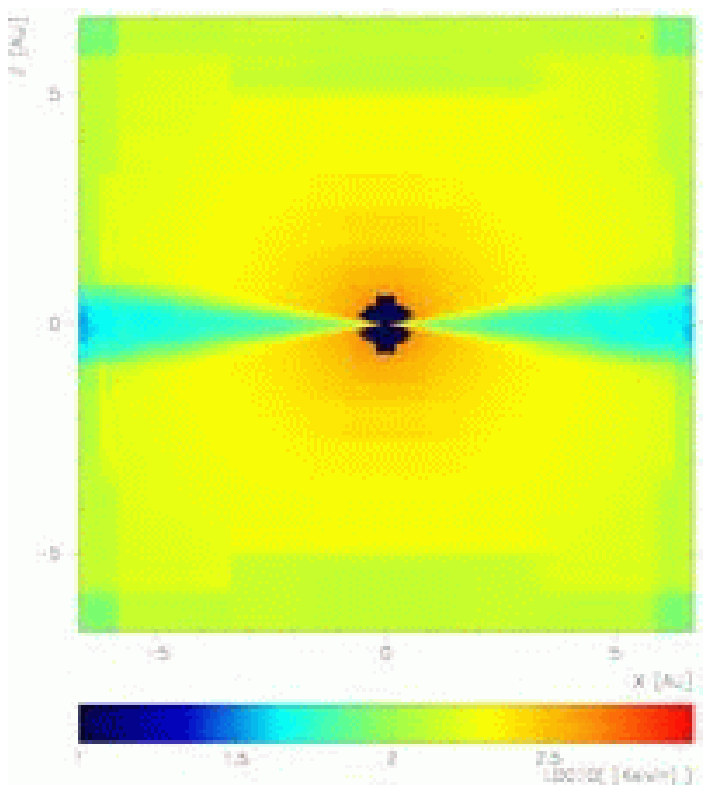}\tabularnewline
\end{tabular}

\vspace{1.5cm}

\begin{tabular}{cc}
\includegraphics[%
  scale=0.4]{fig04c.eps}&
\includegraphics[%
  scale=0.4]{fig04d.eps}\tabularnewline
\end{tabular}

\end{center}

\caption{The density (upper left) and the temperature (upper right)
  maps around a typical star (object 4 in Table~\ref{tab:catalog}),
  are shown for Model II. No material would be seen in a similar plot
  for Model~I on the same scale due to the absence of small scale
  discs.  In the temperature map, the hole around the star was created
  because the temperature of the low density material present in the
  background became more than 1600~K, the dust sublimination
  temperature, and the cells were removed from the computational
  domain. The density at the mid-plane of the disc is plotted as a
  function of the distance from the star (lower left).  The
  temperature at the mid-plane is plotted as a function of the
  distance from the star (lower right). The temperature slope of the
  outer part of the disc (0.3--10 au) is $-0.45$ (dash-dot line)
  which is consistent with that of \citet{whitney:2003a}.  This is
  slightly smaller than the value for an optically thin disc ($-0.4$),
  with an absorption coefficient inversely proportional to wavelength,
  heated by stellar radiation (\citealt{spitzer:1978};
  \citealt{kenyon:1993}; \citealt{chiang:1997}). In the inner part of
  the disc (0.1--0.12~au), the slope quickly increases. A similar
  rapid raise of the slope is also seen in Fig.\,6 of
  \citet{whitney:2003a}. }

\label{fig:diskmodel}

\end{figure*}

The density and the temperature maps around our typical star, object 4
(see Table~\ref{tab:catalog}), are shown in Figure~\ref{fig:diskmodel}
(upper right and left) for Model II. On the same scale, no structure
will be seen in a similar plot for Model I as explained in
Section~\ref{sub:Dust-Model}.  The density map clearly shows that the
cell size increases as the distance from the star (located at the
centre) increases, and also as the distance from the mid-plane of the
disc increases. In the temperature map, the hole around the star is
created because the temperature of the low density material present in
the background reaches 1600~K, the dust sublimination temperature;
therefore, those cells are removed from the computational domain. We
note that temperatures of the inner-most part of discs also can become
greater than 1600~K, and the grid cells in this part of the discs will
be removed. As a consequence, the inner radius may be slightly
larger than an initial value ($R_{\mathrm{i}}=6\, R_*$).

In the lower left of Figure~\ref{fig:diskmodel}, the density at the
mid-plane of the disc is plotted as a function of the distance from
the star. The slope of the density plot is $\sim-15/8$ as it should
be~according to
equations~\ref{eq:disc-density-function}--\ref{eq:scale-height}.
Similarly, the temperature at the mid-plane is plotted as a function
of the distance from the star in the lower right of the same figure.
The slope at the outer part of the disc (0.3--10~au) is $-0.45$ which
is consistent with that of \citet{whitney:2003a}. This is slightly
smaller than the value for an optically thin disc ($-0.4$), with an
absorption coefficient inversely proportional to wavelength, heated by
stellar radiation (c.f., \citealt{spitzer:1978};
\citealt{kenyon:1993}; \citealt{chiang:1997}).  In the inner part of
the disc (0.1--0.12~au), the slope quickly increases.  A similarly rapid
rise of the slope is  seen in the no-accretion-luminosity disc
model of \citet{whitney:2003a} as shown in their Fig.~6.

\subsection{Spectral energy distributions of the cluster \label{sub:SED-cluster}}

\begin{figure}

\begin{center}

\includegraphics[%
  bb=8bp 36bp 560bp 533bp,
  clip,
  scale=0.45]{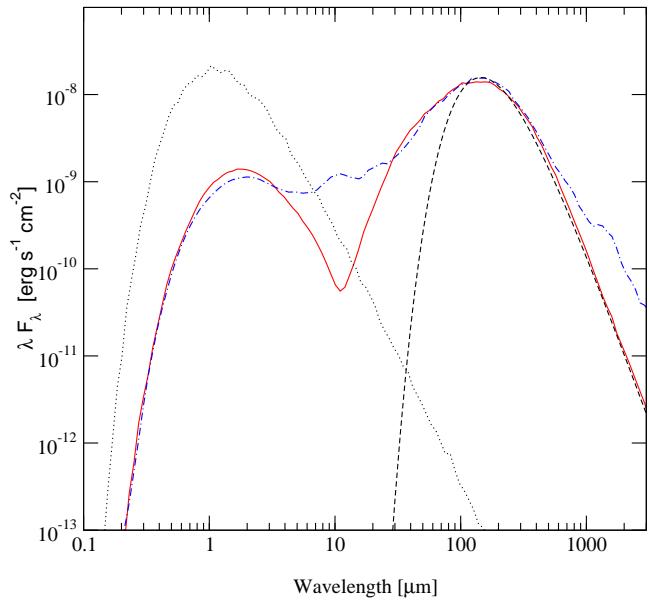}

\end{center}

\caption{The spectral energy distributions (SEDs) from Model I (solid
line) \& Model II (dash-dot line) compared with the 20~K blackbody
radiation curve (dashed line) and the SED of naked (without the ISM
and circumstellar dust) stars and brown dwarfs (dotted line). Both
models  show  peaks around 2~$\mathrm{\mu m}$ and 200~$\mathrm{\mu
m}$. The first peak corresponds to that of stellar emission, and the
latter corresponds to  emission from the ISM dust with $T \sim
20~\mathrm{K}$. The excess emission between  3~$\mathrm{\mu m}$ and
30~$\mathrm{\mu m}$ for Model II is caused by the warm dust in the
accretion discs. The excess emission in the far infrared ($\lambda >
500~\mathrm{\mu m}$) seen in Model II is due to the emission from the
increased amount of cold ($T<20~\mathrm{K}$) dust due to disc
shadowing.}

\label{fig:sed_compare}

\end{figure}

The SEDs from Models I and II are shown in Figure~\ref{fig:sed_compare}
along with the SED of the combined naked stars (the stars without
the ISM and the circumstellar dust) and a 20K blackbody radiation
spectrum. The observer is placed at a
distance of 140~pc (on the $+z$ axis) from the cluster. 
Both models show peaks around $2\,\mathrm{\mu m}$ and $200\,\mathrm{\mu m}$.
The first peak corresponds to that of stellar emission, and the latter
corresponds to the emission from the relatively low temperature ISM
dust. In Model I, the $200\,\mathrm{\mu m}$ peak is prominent, indicating
that the most of the ISM dust has $T\approx20\,\mathrm{K}$. 

The most noticeable differences between the two models are the flux
levels at $\lambda=$~3--30~$\mathrm{\mu m}$ and
$\lambda>500\,\mathrm{\mu m}$.  The excess emission of Model II at
$\lambda=$~3--30~$\mathrm{\mu m}$ is due to the warm dust present in
the accretion discs. This emission arises from the reprocessing of
photospheric emission by the inner discs into the observer's
line-of-sight, and is dominated by a handful of objects (in particular
object numbers 7 and 8, which constitute an ejected binary
system). The excess emission in the far infrared ($\lambda>500\,\mathrm{\mu
  m}$) in Model II is caused by the presence of a larger amount of
cold ($T<20\,\mathrm{K}$) dust. As we can see from
Figure~\ref{fig:temperature-map}, a considerable fraction of the ISM
in Model II is underheated compared to the dust in Model I because the
radiation from the light sources is blocked/shadowed by the accretion
discs. The SED of Model II is very similar to the observed SED of a
young star forming region (ISOSS~J~20298+3559-FIR1) with mass around
120$\mathrm{M_{\sun}}$ presented by \citet{krause:2003}. They showed
that a Planck function with the temperature corresponding to the peak
of the SED ($T=19\,\mathrm{K}$) underestimates the flux levels
observed at $\lambda<60\,\mathrm{\mu m}$ and $\lambda>500\,\mathrm{\mu
  m}$. 

We note that the slope of the SEDs at sub-milimetere wavelengths is
sensitive to the wavelength dependency of the dust absorption
coefficient ($\kappa$). If a shallower wavelength dependency of the
dust opacity is introduced for the discs in Model II, the slope of the
SEDs in the sub-millimetre wavelength range may change
(e.g. \citealt{beckwith:1990}; \citealt{beckwith:1991}).

%
%
% mass estimate section deleted from here.
%
%

\subsection{Spectral energy distributions of stars and brown dwarfs\label{sub:sed-individual}}

Detailed studies of isolated objects have been performed by
\citet{wood:2002b} and \citet{whitney:2003b}, covering the evolution
of the SED from the protostar through to the remnant disc stage.
These models are an excellent point of comparison for our model SEDs,
which are based on the density structures predicted for a 'realistic'
star-forming cloud. The circumstellar material from the SPH
calculation is however significantly more complex than the
axisymmetric, idealized models cited above.

\begin{figure*}

\begin{center}

\includegraphics[%
  clip,
  scale=0.95]{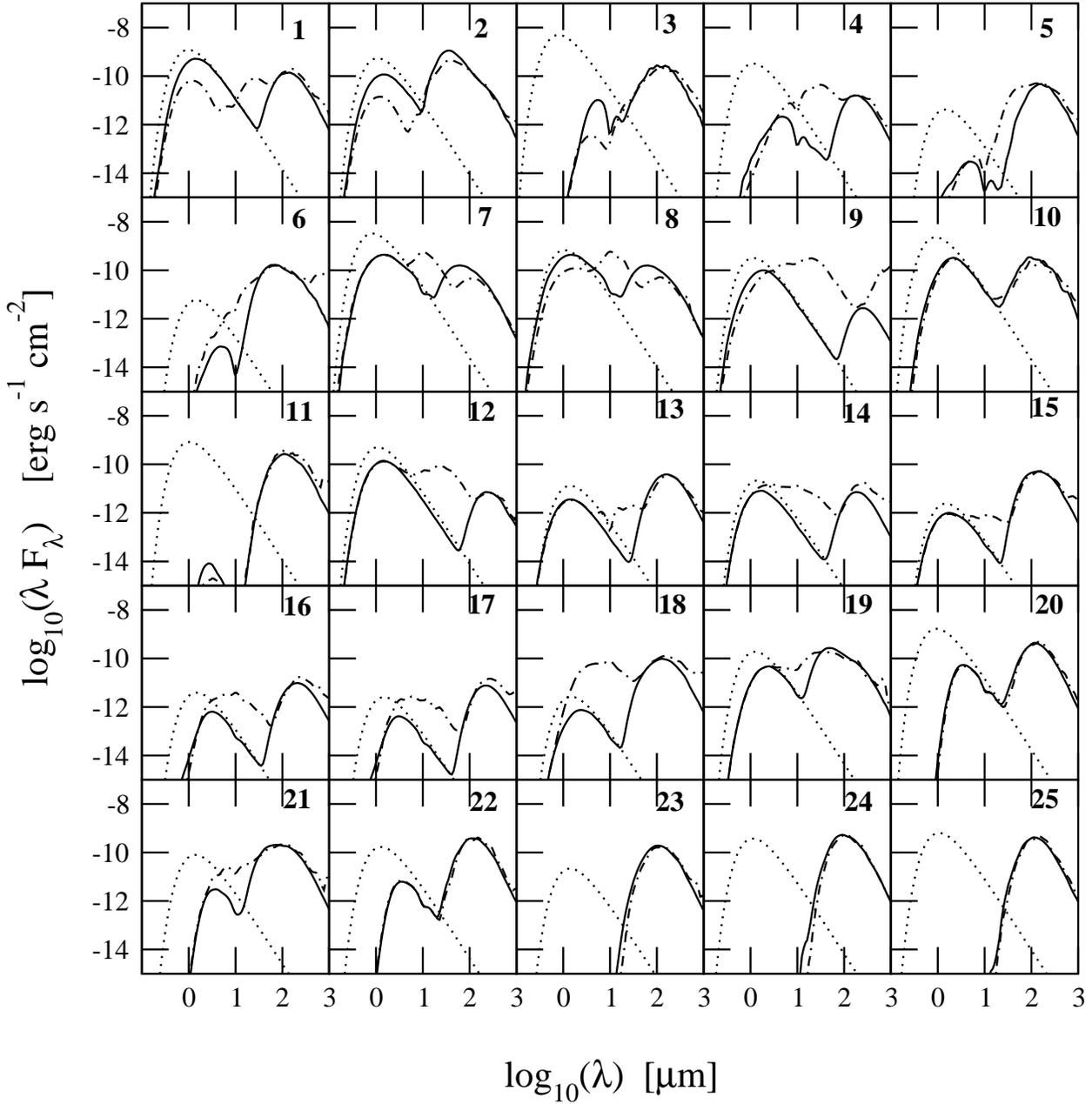}

\end{center}

\caption{Atlas of individual spectral energy distributions (SEDs). The
SEDs from Model I are shown in solid lines and those from Model II in
dash-dot lines. The input source SEDs are presented in dotted lines.
The numbers shown in upper right hand corner of each plot
correspond to the object ID number in Table~\ref{tab:catalog}.  To
compute the SED from each star, the same density and temperature
structures in section~\ref{sub:Density-and-temperature} are used with
some restrictions (see text for detail). As we can see from the SEDs,
the sources are deeply embedded in a cloud, and about the half of them
show little  flux in the optical.  The silicate absorption feature at
10~$\mathrm{\mu m}$ becomes more prominent for objects with higher
extinction. }

\label{fig:sed_atlas01}

\end{figure*}

To calculate the SEDs of individual objects, we have used the same
density and temperature structures as in
Section~\ref{sub:Density-and-temperature} with some restrictions.
Firstly, a cylinder of radius 50~au with its centre passing through an
observer (situated on the $z$-axis) and the centre of an object is
considered.  (The diameter of this cylinder roughly corresponds to a
0.7 arcsec aperture at 140\,pc).  We note that
\citet{whitney:2003b} used 1000~au and 5000~au aperture sizes to
compute colours of protostars, and found that the colours can change
depending on the aperture size adopted. In our calculations, we use a
smaller (50~au) aperture because the stellar densities in the cluster
are such that larger apertures may include significant contributions
from neighbouring objects. Secondly, the dust emission, absorption and
scattering outside of the cylinder are turned off. Thirdly, only the
star under consideration can emit the photons. With these
restrictions, we have performed the Monte Carlo radiative transfer
calculations for each object in Table~\ref{tab:catalog}. The results
are shown in Figure~\ref{fig:sed_atlas01}.  Because of the
restrictions used, some scattered flux contribution from the outside
of the cylinder might be missing in the resulting SEDs, and the
contamination due to the dust emission from the disc of a companion
might be present in the SEDs if a star is in a binary system.  Note that
the binary pairs in the source catalogue (Table~\ref{tab:catalog}) are
3--10, 7--8, 20--22, 44--42, 26--40, 39--41 and 45--38. Readers are
referred to table~2 of \citet{bate:2003a} for binary parameters
(separations, eccentricities and so on). As we can see from the panels of
Figure~\ref{fig:sed_atlas01}, the sources are deeply embedded in the
cloud, and about half of them show little flux in the optical. The
SEDs are very similar to those of the Class~0 model in
\citet{whitney:2003b}. The silicate absorption feature at
10~$\mathrm{\mu m}$ becomes more prominent for objects with higher
extinction.

We have computed $K$ band magnitude, the flux at 8~$\mathrm{\mu m}$
($F_{8}$) and 24~$\mu m$ ($F_{24}$) based on the SEDs, and extinction
($A_{V}$) by integrating the opacities between each object and an
observer.  The results are placed in Table~\ref{tab:catalog}.  For
Model II, the inclinations of the discs with respect to an observer on
$+z$ axis (that used in the calculation of the global SEDs) are also
listed in the same table. The number of objects brighter than $K=20$
are 26 and 30 objects for Model I and Model II respectively. 

\addtocounter{figure}{-1}

\begin{figure*}

\begin{center}

\includegraphics[%
  clip,
  scale=0.95]{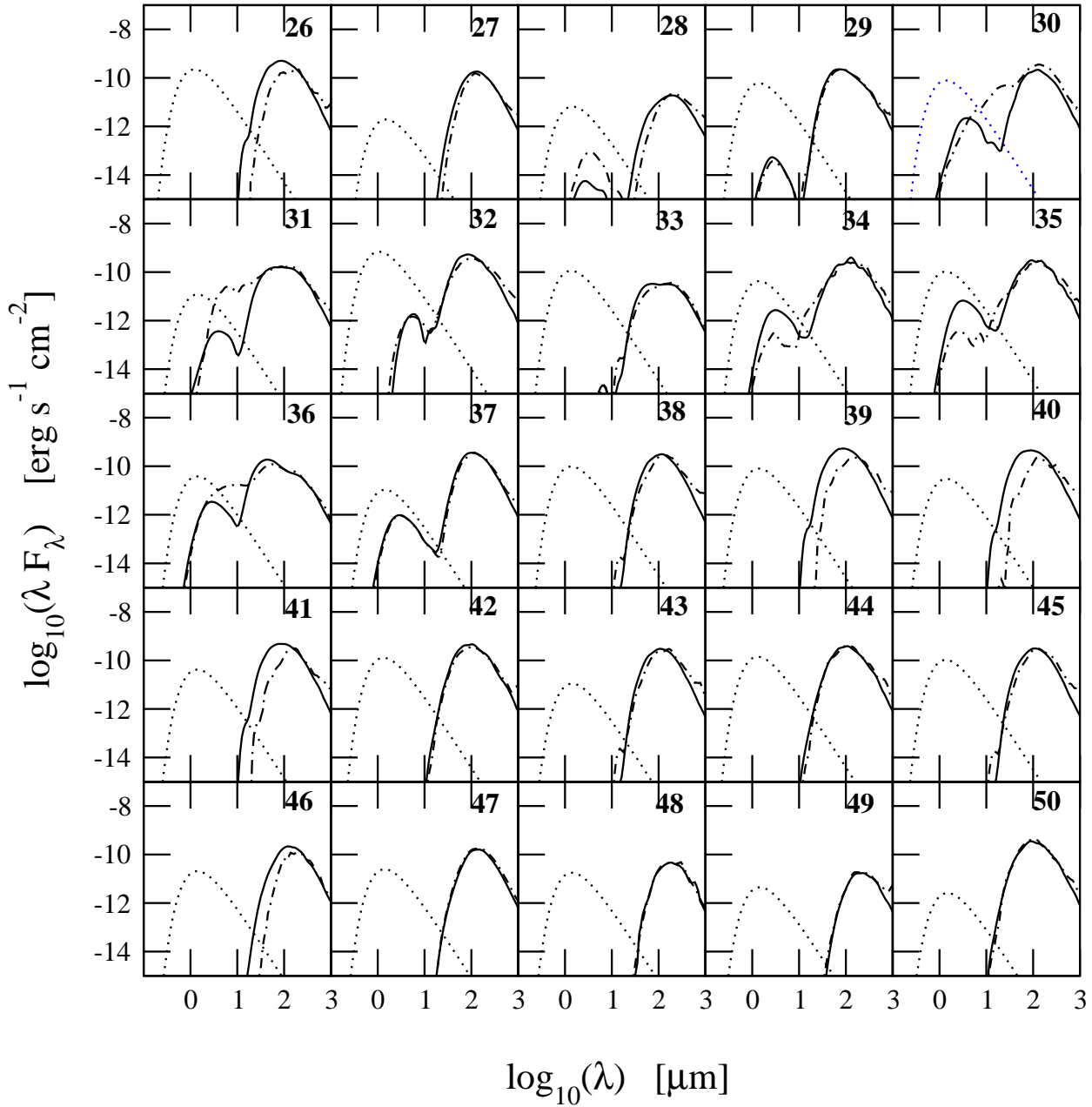}

\end{center}
\caption{Concluded.}

\label{fig:sed_atlas02}

\end{figure*}

Using the flux between 2.2--10.2~$\mu$m, the SEDs in
Figure~\ref{fig:sed_atlas01} are classified based on the spectral
index $\left(a\right)$ as defined by \citet*{wilking:1989}: \[
a=\frac{d\log\left(\lambda F_{\lambda}\right)}{d\log\lambda}.\] We
have used the monochromatic fluxes at \emph{K} (2.2\,$\mathrm{\mu
  m}$), \emph{L} (3.6\,$\mathrm{\mu m}$), \emph{M} (4.8\,$\mathrm{\mu
  m}$), \emph{N} (10.2\,$\mathrm{\mu m}$) to compute the values of $a$
via least-squares fits (e.g.  \citealt{greene:1994};
\citealt{haisch:2001}). The classification scheme of
\citet{greene:1994} is adopted in our analysis. The sources with
$a>0.3$ are classified as `Class~I', $-0.3<a\leq0.3$ as `flat
spectrum', $-1.6<a\leq-0.3$ as `Class~II', and $a<-1.6$ as `Class~III'
YSOs. In addition to this scheme, we classify sources to be Class~0
(\citealt{andre:1993}) if the ratio of $\log\lambda F_{\lambda}$
values at $\lambda=160\,\mathrm{\mu m}$ (\emph{SST} MIPS) and at
$\lambda=2.2\,\mathrm{\mu m}$ ($K$ band) is greater than 3.0 (c.f.,
Fig.~3 of \citealt{whitney:2003b}). The results are placed in
Table~\ref{tab:catalog} for both models. Out of 50 objects, there are
27 Class~0, no Class~I, no flat spectrum, 11 Class~II and 12 Class~III
objects for Model I, and there are 28 Class~0, 9 Class~I, 4 flat
spectrum, 6 Class~II, and 3 Class~III objects for Model II. Note that
even though all objects are $<0.07$ Myrs old, there is a mixture of
Class 0--III objects. Thus, the class of an object does not necessarily
relate to its evolutionary stage.  \citet{whitney:2003b} also pointed
out the degeneracy problem of the SEDs from Class~I and Class~II
objects, and that their colours look very similar to each other for
some combinations of disc inclinations and luminosities.

\begin{figure*}
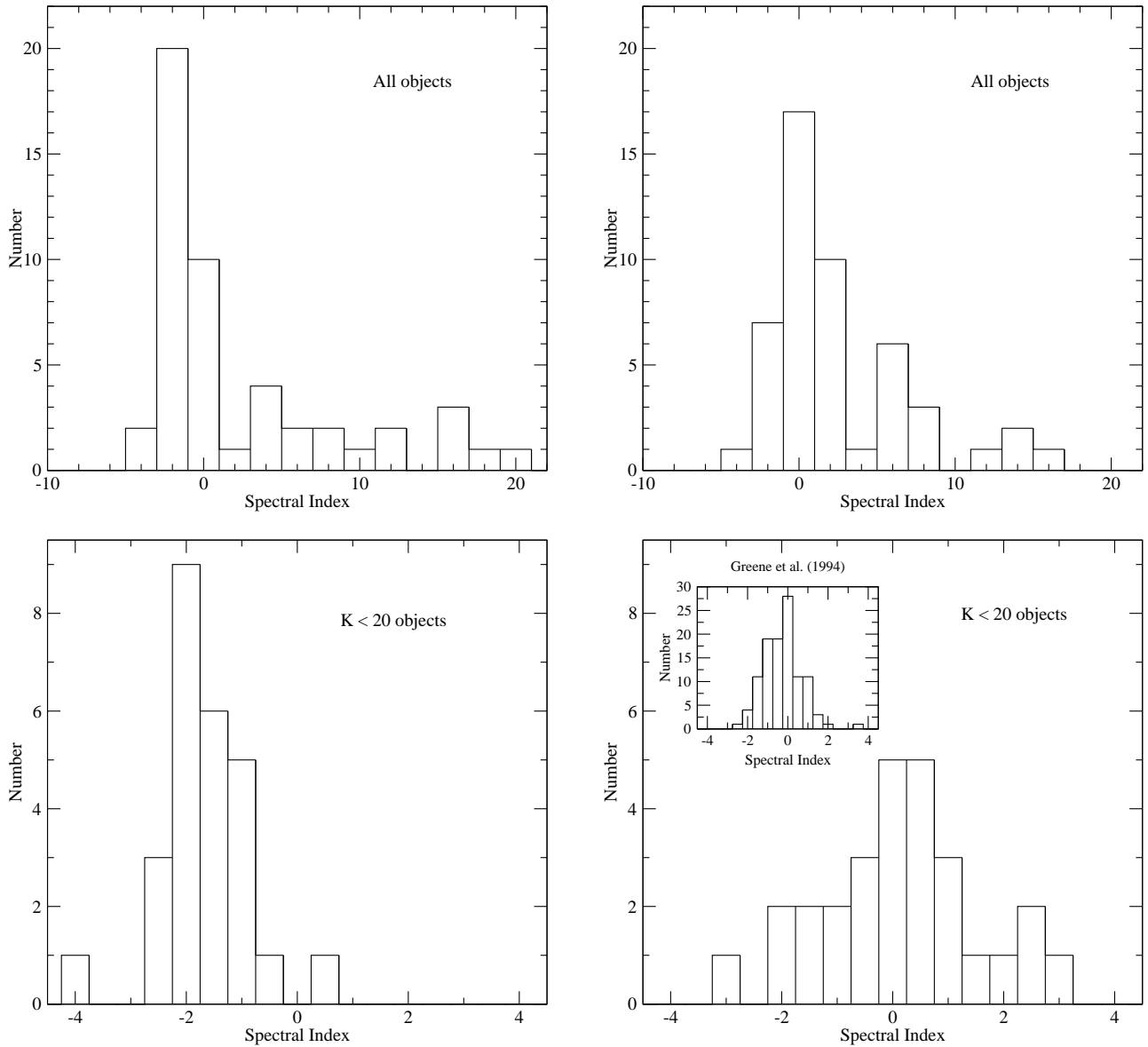


\begin{center}

\vspace{0.5cm}

\begin{tabular}{cc}
\includegraphics[%
  bb=8bp 36bp 543bp 533bp,
  clip,
  scale=0.45]{fig07a.eps}&
\includegraphics[%
  bb=8bp 36bp 543bp 533bp,
  clip,
  scale=0.45]{fig07b.eps}\tabularnewline
\includegraphics[%
  bb=8bp 36bp 543bp 533bp,
  clip,
  scale=0.45]{fig07c.eps}&
\includegraphics[%
  bb=8bp 36bp 543bp 533bp,
  clip,
  scale=0.45]{fig07d.eps}\tabularnewline
\end{tabular}

\end{center}

\caption{The distributions of the spectral indices for Model I (left)
and Model II (right) computed from the flux levels at $K$, $L$, $M$
and $N$ bands ($2.2-10.2~\mathrm{\mu m}$) in the SEDs shown in
Figure~\ref{fig:sed_atlas01}. While the
upper plots include all 50 objects in our source catalogue
(Table~\ref{tab:catalog}), the lower plots include only the objects
with $K < 20$.  The distribution for Model II with $K<20$ (lower
right) is very similar to that from the observation of the
$\rho$~Ophiuchus cloud by \citet{greene:1994} shown in the inset. }

\label{fig:spectral_index}

\end{figure*}

Figure~\ref{fig:spectral_index} shows the distribution of the spectral
index values. The upper row in the figure shows the histograms of the
index values from all 50 objects while the lower part shows those
created from 'observable' ($K<20$) objects. The histograms based on
all objects peak around $a=0$ for Model I and II, and the
distributions are skewed. On the other hand, the number of the objects
with $K<20$ for Model II look more evenly distributed around $a=0$
(lower right in Figure~\ref{fig:spectral_index}). The latter is very
similar to the spectral index distribution of the $\rho$~Ophiuchus
cloud presented in Fig.~3 of \citet{greene:1994}, although we have to
include fainter objects in order to have sufficient indices to make a
comparison.

\subsection{Multi-band far-infrared images\label{sub:Multi-Band-Far-Infrared-Images}}

\begin{figure*}

\begin{center}

\begin{tabular}{ccc}
\includegraphics[%
  scale=0.7,
  angle=270]{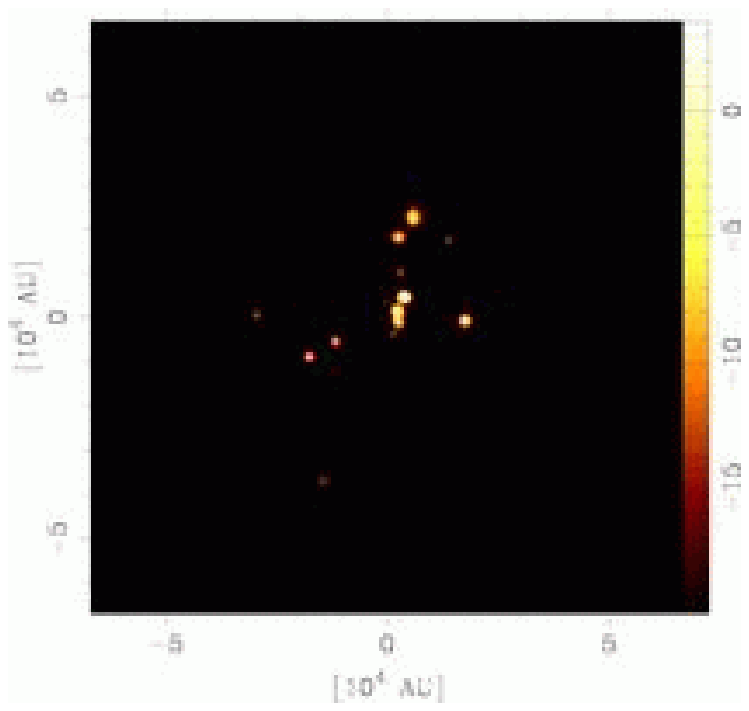}&
\includegraphics[%
  scale=0.7,
  angle=270]{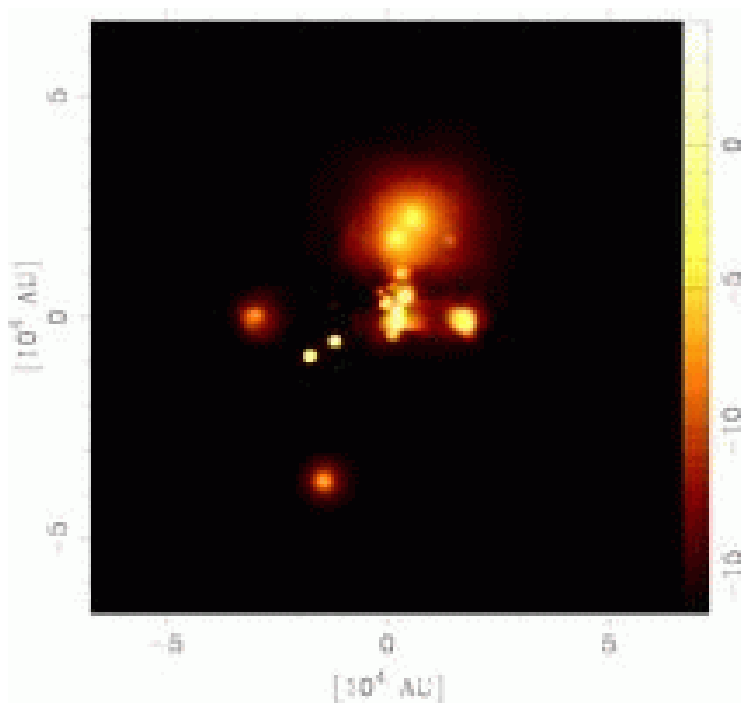}&
\includegraphics[%
  scale=0.7,
  angle=270]{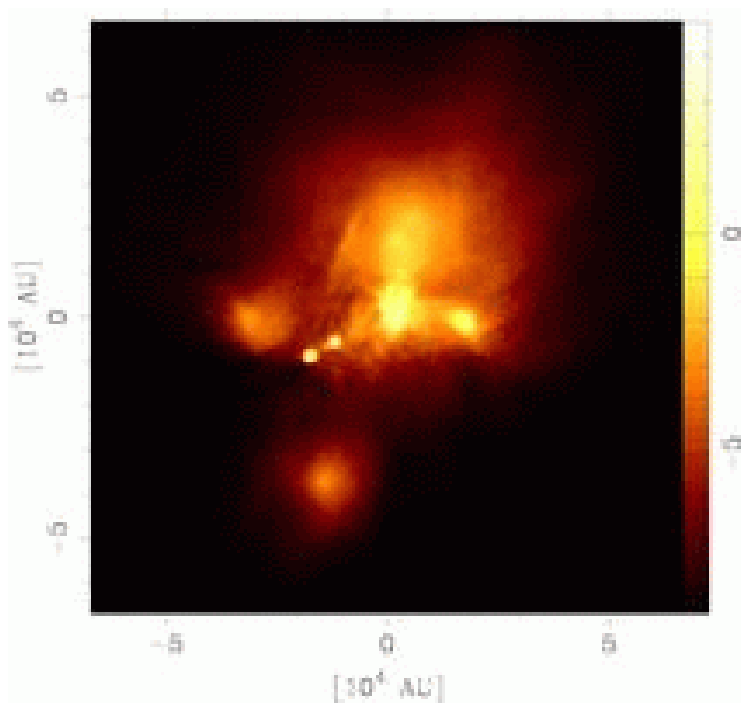}\tabularnewline
\includegraphics[%
  scale=0.7,
  angle=270]{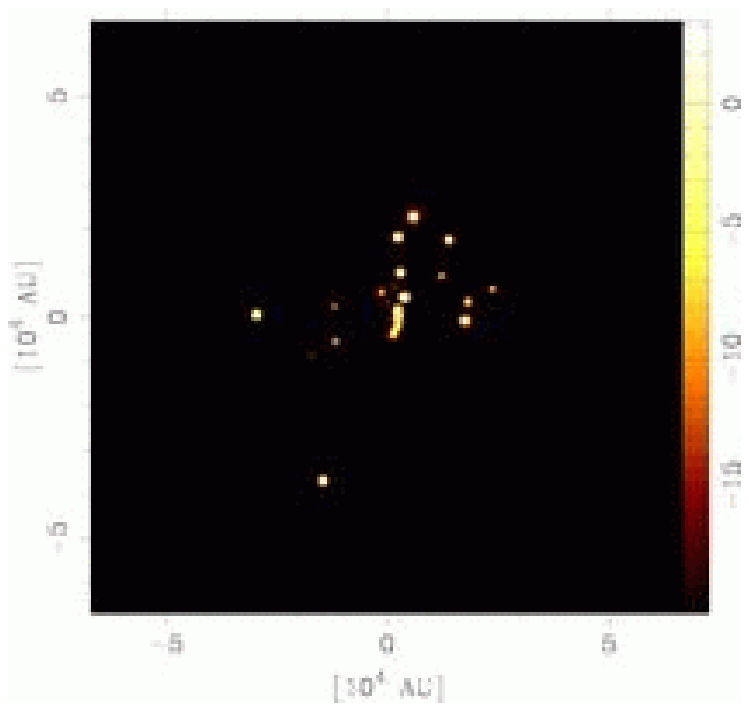}&
\includegraphics[%
  scale=0.7,
  angle=270]{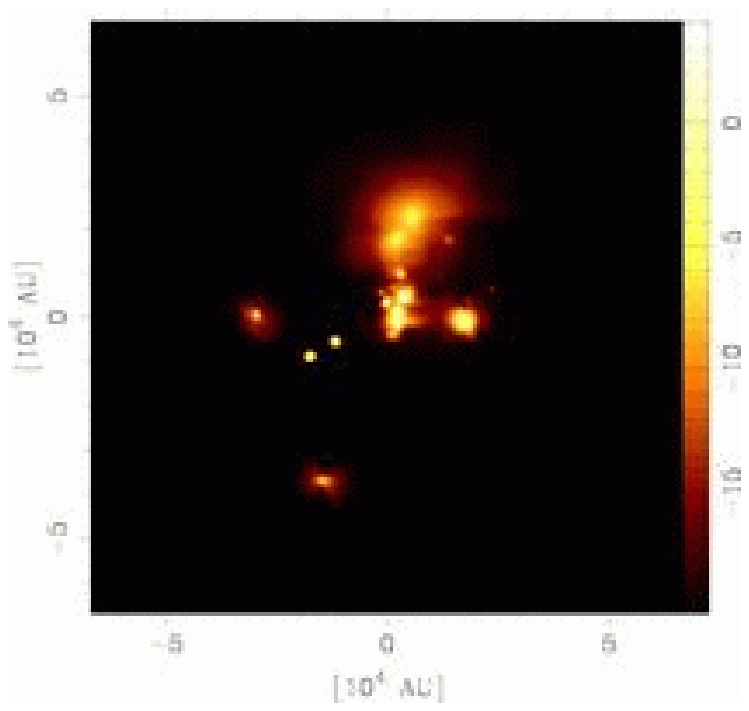}&
\includegraphics[%
  scale=0.7,
  angle=270]{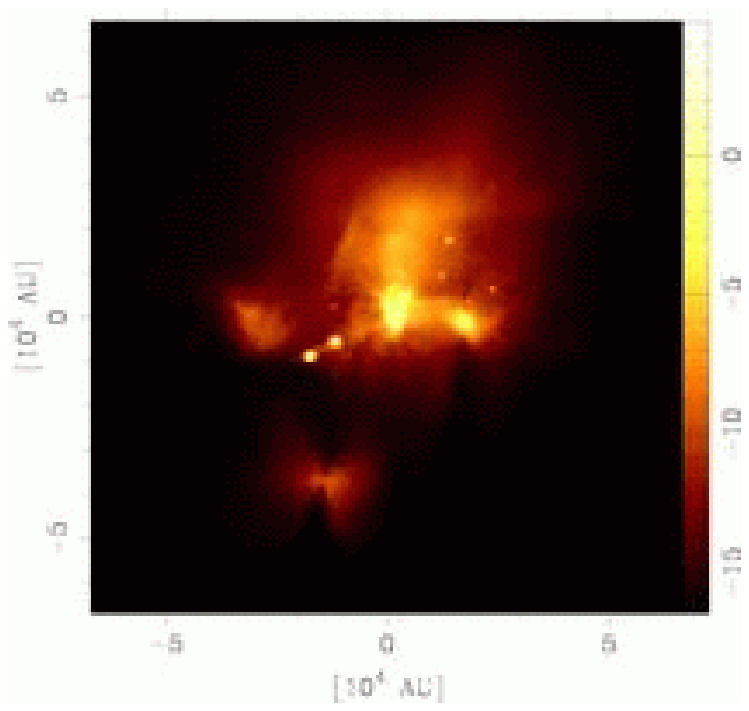}\tabularnewline
\end{tabular}

\end{center}

\caption{Top: Idealised images for Model I at \em SST \em MIPS 24, 70,
and 160~$\mathrm{\mu m}$ bands from left to right.  Bottom: Same as
the top images, but for Model II. The units used for the colour scale
in the figures are log of $\mathrm{MJy~sr^{-1}}$.  The cluster is
placed at 140~pc from the observer, and the corresponding angular size
of the images is  $15.8\times15.8~\mathrm{arcmin^{2}}$. Most of the
objects appear brighter in the Model II images due to the warm dust
emission from the circumstellar discs.  The presence of the disc
influences the morphology of the dust emission in the 70 and 160
$\mathrm{\mu m}$ images. The butterfly structures caused by the almost
edge-on disc can be seen in the lower part of the 70 and 160
$\mathrm{\mu m}$ images for Model II.} 

\label{fig:mips-images}

\vspace{1.5cm}

\end{figure*}

Figure~\ref{fig:mips-images} shows simulated images for the \emph{SST}
MIPS bands with central wavelengths ($\lambda_{c}$) at 24, 70 and
160~$\mathrm{\mathrm{\mu m}}$. These are \emph{idealised} images in
which the resolutions are limited only by the number of pixels used in
simulations ($500^{2}$ pixels), and are not degraded to the
diffraction limit of the \emph{SST}. The observer is placed at a
distance of 140~pc (on the $+z$ axis) from the cluster. At this
distance, the images subtend about
$15.8\times15.8\,\mathrm{arcmin}^{2}$.  As predicted from the model
SEDs (Figure~\ref{fig:sed_compare}), most of the objects appear
brighter for Model II in the $24\,\mathrm{\mathrm{\mu m}}$ image
because of the warm dust emission from the circumstellar discs.
Additional images for Model II with the observers on the $+x$ and $+y$
axis are also computed to create 3-colour, red
(160~$\mathrm{\mathrm{\mu m}}$), green (70~$\mathrm{\mathrm{\mu m}}$)
and blue (24~$\mathrm{\mathrm{\mu m}}$), composite images. The results
are shown in the third row of Figure~\ref{fig:ColumnDensity}.

Although there is little effect of the circumstellar discs on the
predicted SEDs around 70 and 160~$\mathrm{\mu m}$, the presence of the
discs influences the morphology of the dust emission. In the lower
part of the 70 and 160~$\mathrm{\mu m}$ images of Model II, the
`butterfly' structure is caused by an almost edge-on disc. The
structure resembles the near infrared images of the edge-on discs
around T~Tauri stars (e.g. HK~Tau~C by \citealt{stapelfeldt:1998};
HH~30~IRS by \citealt{burrows:1996}) but while these butterflies are
formed by scattering, those presented here are the result of thermal
processes.  Moreover, the scale size of the butterfly structure in the
70 and 160~$\mathrm{\mu m}$ images of Model II is much larger
($\sim10^{4}$~au) than that seen in the near infrared observations. The
typical radius of the circumstellar discs around T~Tauri stars based
on the near-infrared morphology is order of $\sim10^{2}$~au (e.g. see
\citealt{burrows:1996}; \citealt{lucas:1998};
\citealt{stapelfeldt:1998}).  Interestingly, a recent observation of
the T~Tauri star ASR~41 in NGC~1333 by \citet{hodapp:2004} showed that
the dark band in the reflection nebula around the star can be traced
out to $\approx3000$~au from the centre. Based on their radiative
transfer models, they concluded that the large appearance of ASR~41 is
probably caused by the shadow of a much smaller disc ($\sim10^{2}$ au)
being projected into the surrounding dusty cloud.

\begin{figure*}

\begin{center}

\begin{tabular}{ccc}
\includegraphics[%
  scale=0.7,
  angle=270]{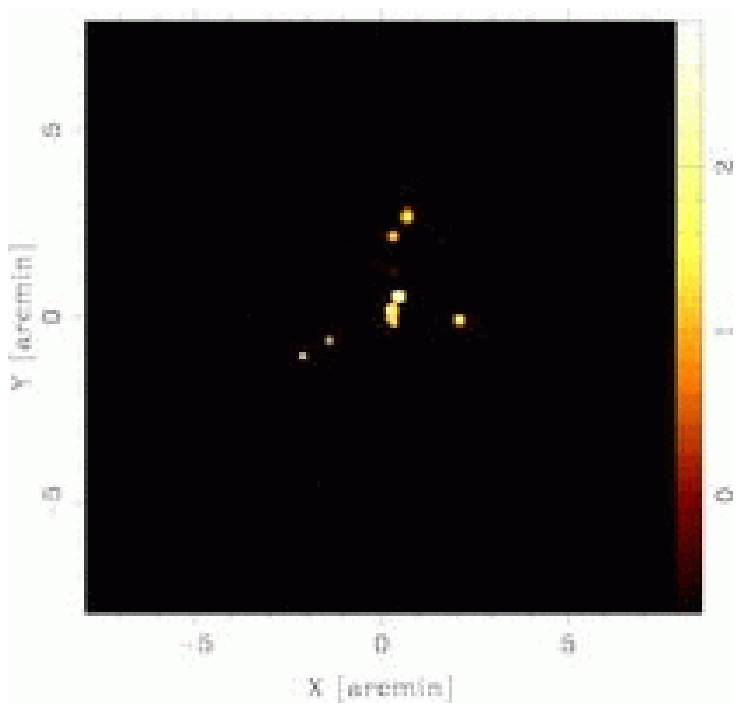}&
\includegraphics[%
  scale=0.7,
  angle=270]{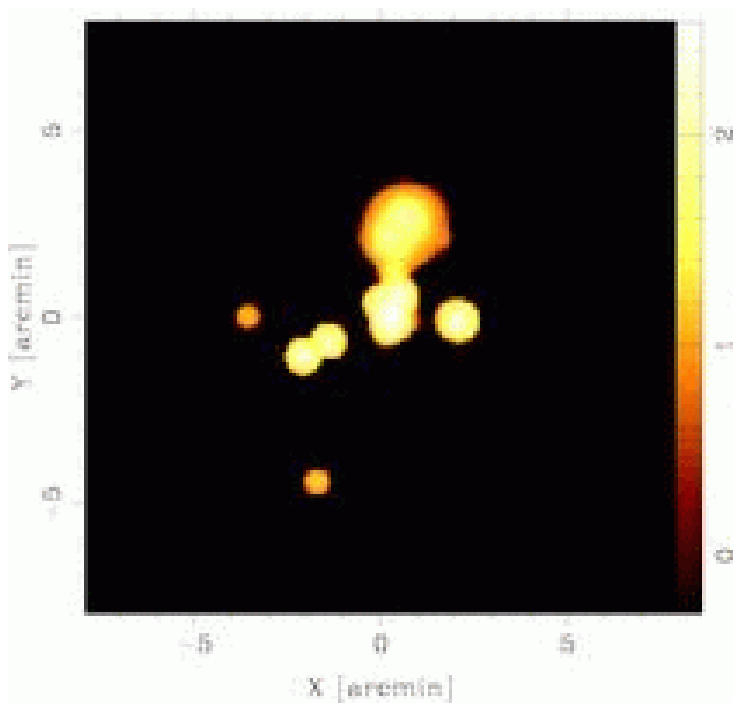}&
\includegraphics[%
  scale=0.7,
  angle=270]{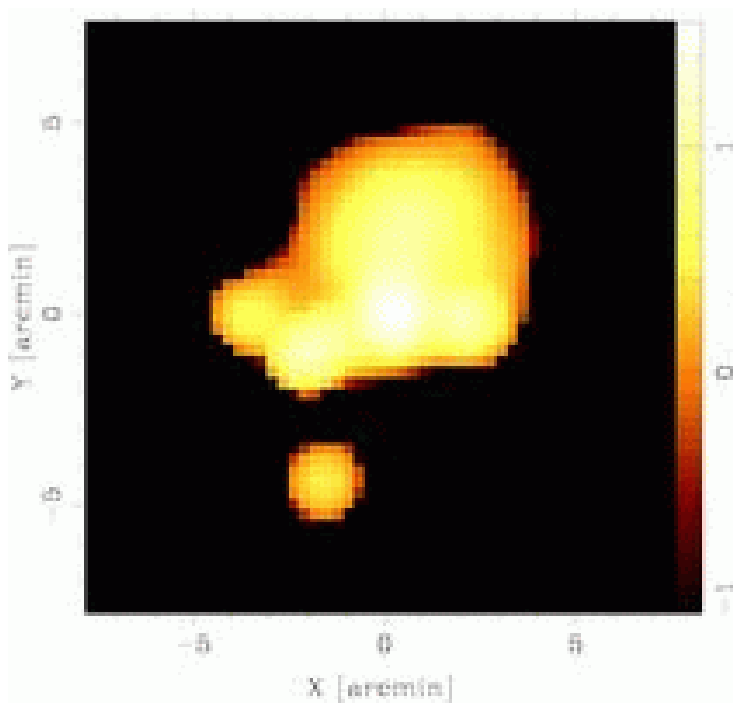}\tabularnewline
\includegraphics[%
  scale=0.7,
  angle=270]{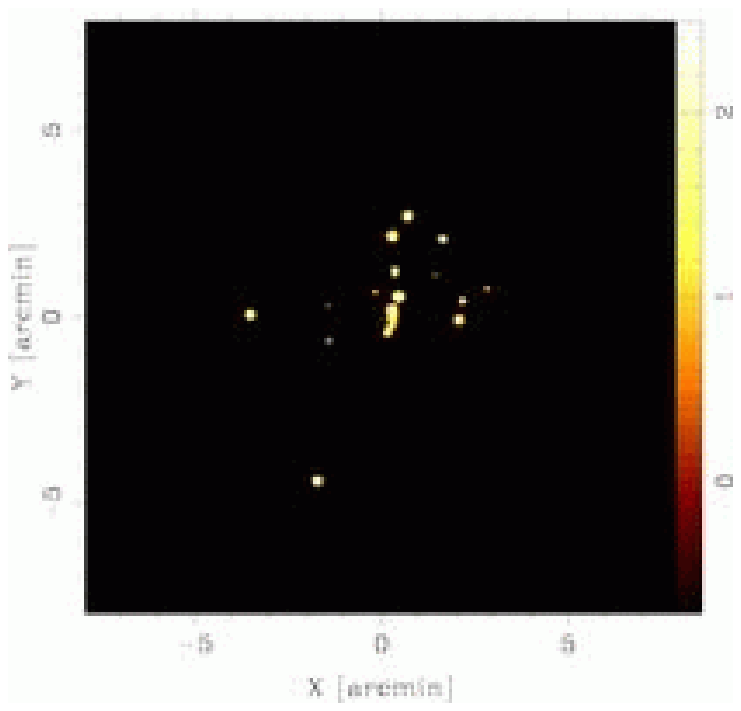}&
\includegraphics[%
  scale=0.7,
  angle=270]{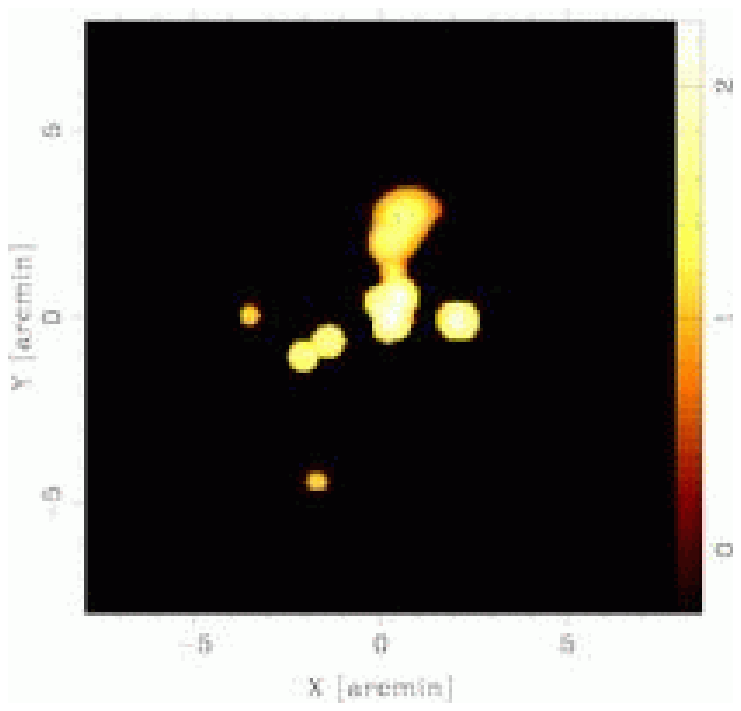}&
\includegraphics[%
  scale=0.7,
  angle=270]{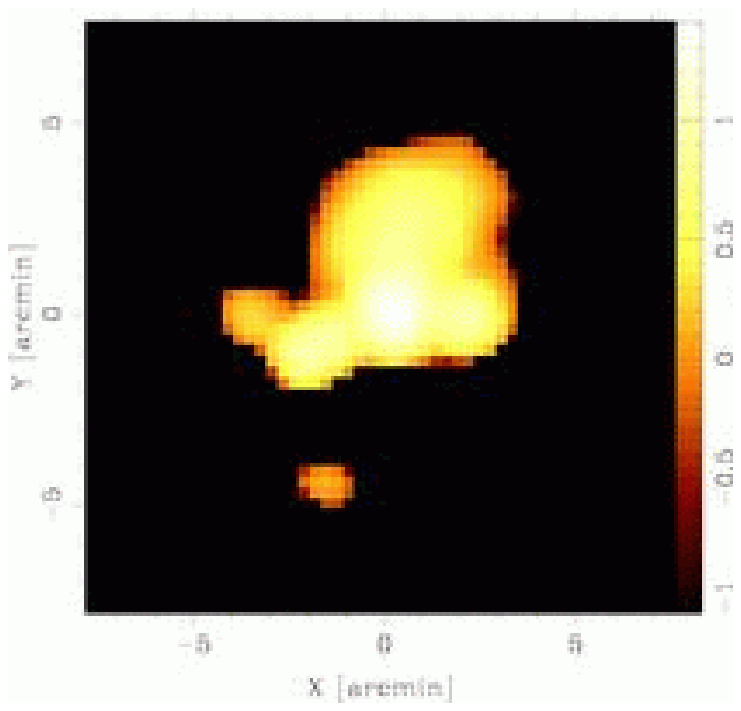}\tabularnewline
\end{tabular}

\end{center}

\caption{Top: Simulated  \em SST \em MIPS images for Model I at  24,
  70, and 160~$\mathrm{\mu m}$ bands from left to right.  Bottom: Same
  as the top images, but for Model II. As in
  Figure~\ref{fig:mips-images}, the colour scale values are in log of
  $\mathrm{MJy~sr^{-1}}$. The images are degraded to the diffraction
  limits, 7.6, 22.1 and 50.4~arcsec for 24, 70, and 160~$\mathrm{\mu
    m}$ images respectively, by convolving them with a Gaussian
  filter. Further, the image pixels are combined to match the
  instrument resolutions of MIPS (2.55, 4.99 and 16.0~arcsec for 24,
  70, and 160~$\mathrm{\mu m}$ images respectively). The lower flux
  cut (the minimum value of the flux scale in each image)
  approximately corresponds to the MIPS limiting fluxes (0.2, 0.5 and
  0.1~MJy\,$^{-1}$ for 24, 70 and 160~$\mathrm{\mu m}$ bands
  respectively). No clear distinction can be made between the two
  models at 70 and 160~$\mathrm{\mu m}$.  The dust emission from the
  circumstellar disc affects the number of observable stars in the
  $24~\mathrm{\mu m}$ images.}

\label{fig:mips-images-limit}

\end{figure*}

To construct simulated images for actual observations by the
\emph{SST}, the images shown in Figure~\ref{fig:mips-images} are
degraded to the diffraction limits of an 85\,cm telescope; 7.6, 22.1
and 50.4~arcsec for 24, 70, and 160~$\mathrm{\mu m}$ images
respectively, by convolving them with a Gaussian filter, and then the
image pixels are binned-up to match the pixel scale of MIPS (2.55,
4.99 and 16.0~arcsec for 24, 70, and 160~$\mathrm{\mu m}$ images
respectively). The results are shown in
Figure~\ref{fig:mips-images-limit}. The lower flux cut (the minimum
value of the flux scale in each image) approximately corresponds to
the MIPS limiting flux for each band (0.2, 0.5 and 0.1~MJy\,sr$^{-1}$
for 24, 70 and 160~$\mathrm{\mu m}$ bands respectively).  There is no
significant change seen in the 24~$\mathrm{\mu m}$ images from the
previous images while the 70, and 160~$\mathrm{\mu m}$ images are
clearly degraded in resolution and sensitivity. No clear distinction
can be made between the two models (with or without the small scale
discs) from the 70~$\mathrm{\mu m}$ images. The major structures in
the 160~$\mathrm{\mu m}$ images appear to be the same in both models,
but the emission from the isolated small structure in the lower half
of the images is much weaker in Model II. According to the simulated
24~$\mathrm{\mu m}$ images, 8 and 18 (out of 50) objects are detected
above the flux limit for Models I and II, respectively.  The simulated
images at 70 and 160~$\mathrm{\mu m}$ show cloud structures with
surface brightnesses that are 10--100 times the \em{SST}\em\,detection limits. 

\subsection{HKL and SST colour-colour diagrams\label{sub:UKIRT-and-SST}}

\begin{figure*}
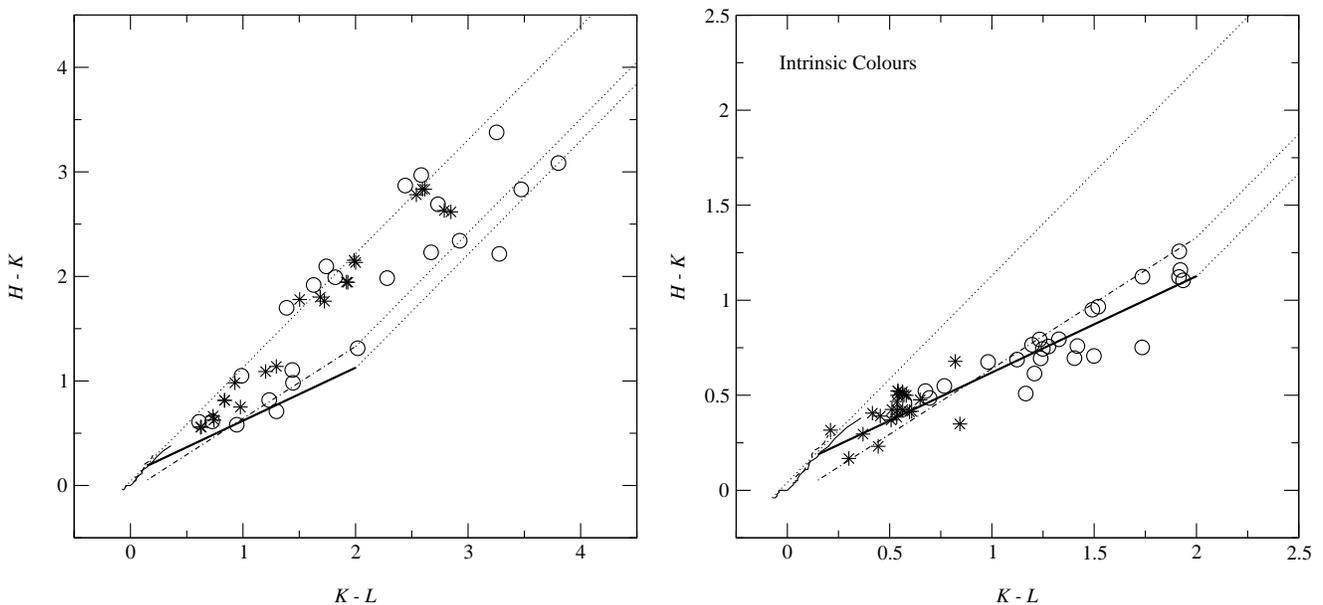


\begin{center}

\begin{tabular}{cc}
\includegraphics[%
  bb=15bp 38bp 543bp 534bp,
  clip,
  scale=0.45]{fig10a.eps}&
\includegraphics[%
  clip,
  scale=0.45]{fig10b.eps}\tabularnewline
\end{tabular}

\end{center}

\caption{The \em H-K \em vs. \em K-L \em colour-colour diagrams for
  the discless objects from Model I (stars) and the disc objects from
  Model II (circles). The left plot shows the colours of objects (with
  $K < 20$) computed from the SEDs given in
  Figure~\ref{fig:sed_atlas01}, and the right plot shows the
  intrinsic (de-reddened) colours of the same objects. The loci of \em
  intrinsic colours \em for the main-sequence stars (solid line), the
  giants (dashed line), and the classical T~Tauri stars (dash-dot
  line) from \citet*{meyer:1997} are also shown along with the
  reddening vectors (dotted lines) extending to the upper right from
  the edges of the loci. The least-squares fit of the intrinsic colours
  of the disc objects (Model~II) are shown in thick solid line (right
  plot).}

\label{fig:color-color-ukirt}

\end{figure*}

\begin{figure*}
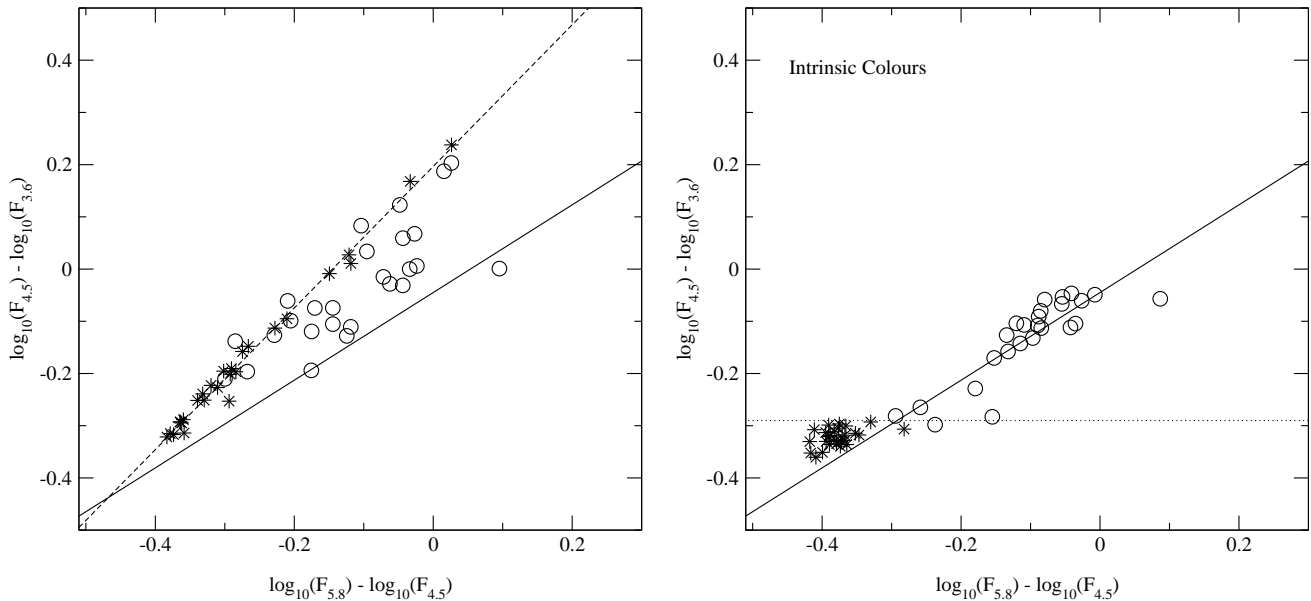


\begin{center}

\begin{tabular}{cc}
\includegraphics[%
  clip,
  scale=0.45]{fig11a.eps}&
\includegraphics[%
  clip,
  scale=0.45]{fig11b.eps}\tabularnewline
\end{tabular}

\end{center}

\caption{The simulated \em SST \em (IRAC) colour-colour diagrams of
  the discless objects from Model~I (stars) and the disc objects from
  Model~II (circles). The $\log_{10}$ of the fluxes at 3.6, 4.5 and
  5.8~$\mathrm{\mu m}$ are used here instead of magnitudes. The left
  plot shows the colours of objects (with $F_{8}
  > 10~\mathrm{\mu Jy}$) computed from the SEDs given in
  Figure~\ref{fig:sed_atlas01}, along with the least-squares fit line
  of discless objects (dashed line).  About 85 per cent of the disc
  objects are located to the right of this line.  The right plot shows
  the \em intrinsic \em (de-reddened) colours of the same objects
  along with the least-squares fit (solid line) of the disc objects.
  The fit (solid line) is also shown in the left plot. While most of
  the disc objects are located above $\log_{10} F_{5.8} - \log_{10}
  F_{4.5} = -0.29$ (dotted line) and along the fit line, the discless
  objects are located below $\log_{10} F_{5.8} - \log_{10} F_{4.5} =
  -0.29$ and along the same fit line.}

\label{fig:color-color-irac}

\end{figure*}

Using the flux levels measured in the SEDs shown in
Figure~\ref{fig:sed_atlas01}, we have constructed simulated
\emph{HKL} and \emph{SST} IRAC (3.6, 4.5 and 5.8~$\mathrm{\mu m}$)
colour-colour diagrams, and placed the results in
Figures~\ref{fig:color-color-ukirt} and \ref{fig:color-color-irac}. We
have also computed the intrinsic (de-reddened) colours of the objects
in Model I and Model II by computing the SEDs without the ISM
absorption (i.e. the opacity of the foreground dust is set to zero).
The results are shown in the same figure for a comparison. For the
\emph{HKL} colour-colour diagram, only the objects with $K<20$ have
been chosen, in order to simulate a deep photometric observation. On the other
hand, the objects with a flux greater than $10\,\mathrm{\mu Jy}$ (at
$8\,\mathrm{\mu m}$) are used to construct the \emph{SST} IRAC
colour-colour diagram.

As we can see from the left-hand plot in Figure~\ref{fig:color-color-ukirt},
the objects with circumstellar discs (Model II) are not well separated
from those without discs (Model I), although the disc objects tend
to be slightly redder than the discless objects. This can be understood
from the SEDs presented in Figure~\ref{fig:sed_compare} which shows
no significant difference between the flux levels of Model I and Model
II in the 1--3~$\mathrm{\mu m}$ wavelength range. The intrinsic
\emph{HKL} colours are less scattered on the diagram. The disc objects
seem to lie along the locus of classical T~Tauri stars given by \citet*{meyer:1997}:
\[
\left(H-K\right)=a\,\left(K-L\right)+b\]
 where $a=0.69\left(\pm0.05\right)$ and $b=-0.05\left(\pm0.04\right)$.
The least-squares fit of our disc object colours gives $a=0.51\left(\pm0.27\right)$
and $b=-0.11\left(\pm0.37\right)$ which are in very good agreement
with the locus of \citet{meyer:1997}. This also indicates our disc
SED models are reasonable. 

The disc objects and discless objects are well separated in the colour-colour
diagram of the \emph{SST} IRAC bands (in the left plot of Figure~\ref{fig:color-color-irac}).
Again, this can be roughly explained from the SEDs in Figure~\ref{fig:sed_compare}.
The SEDs from Model I and Model II start separating from each other
around $\lambda=\mathrm{3\,\mu m}$, and the difference between the
flux levels increases as the wavelength increases until it reaches
about 10~$\mathrm{\mu m}$. Moreover, the flux level of Model I decreases
as a function of the wavelength and vice versa for that of Model II,
in this wavelength range (3--10~$\mathrm{\mu m}$). The least-squares
fit of the discless objects in the left plot of Figure~\ref{fig:color-color-irac}
gives: \[
\left(\log_{10}F_{4.5}-\log_{10}F_{3.6}\right)=a\,\left(\log_{10}F_{5.8}-\log_{10}F_{4.5}\right)+b\]
 where $a=1.36\left(\pm0.14\right)$ and $b=0.20\left(\pm0.04\right)$.
About 85 per cent of the disc objects are located to the right of
this fit line. 

The intrinsic (de-reddened) colours of the objects are computed in the
same way as was done for the \emph{HKL} colours, and the results are
placed in the right-hand plot of Figure~\ref{fig:color-color-irac}.
The least-squares fit of the disc objects is also shown in the same
plot (and also in the left-hand plot). The slope and the intercept of
the line are $a=0.78\left(\pm0.50\right)$ and
$b=-0.04\left(\pm0.05\right)$ respectively. While most of the disc
objects are located above
$\left(\log_{10}F_{5.8}-\log_{10}F_{4.5}\right)=-0.29$ and along the
fit line, the discless objects are located below
$\left(\log_{10}F_{5.8}-\log_{10}F_{4.5}\right)=-0.29$ and along the
same fit line.

Observers (e.g. \citealt{aspin:1994}; \citealt{lada:2000}; \citealt{muench:2001})
often use the \emph{JHK} or \emph{HKL} colour-colour diagrams (either
\emph{J-H} vs \emph{H-K} or \emph{H-K} vs \emph{K-L}) to find candidates
for young stars or brown dwarfs with accretion discs. As we have seen
in this section, a better diagnostic for identifying objects with
accretion discs is to use mid-infrared colours (e.g. \emph{SST} IRAC
bands) instead of the near-infrared colours. This may not be true
for a more massive cluster since the massive stars could produce a
larger fraction of high temperature dust.



\section{Conclusions \label{sec:Conclusion}}

We have presented three-dimensional Monte Carlo radiative transfer
models of a very young low-mass stellar cluster with multiple light
sources (23 stars and 27 brown dwarfs). The density structure and the
stellar distributions from the large-scale SPH simulation of
\citet{bate:2003a} were mapped onto our radiative transfer grid
without loss of resolution using an AMR grid. The temperature of the
ISM and the circumstellar dust was computed using the Monte Carlo
radiative equilibrium method of \citet{lucy:1999b}. The results have
been used to compute the SEDs, the far-infrared \emph{SST} (MIPS bands)
images, and the colour-colour diagrams of this cluster.

We find that the presence of circumstellar discs on scales less than
10~au (Model II) influences the morphology of the temperature
structure of the cluster (Figure~\ref{fig:temperature-map}), and can
affect the temperature of the dust on large scales (up to a several
$10^{4}$~au).  The dust shadowed by accretion discs has a lower
temperature than the model without the discs (Model I). The radiation
coming from within a few au of the light sources is
anisotropic/bipolar because of the circumstellar discs.

The cluster SEDs (Figure~\ref{fig:sed_compare}) from Models I and
II both show peaks around $2\,\mathrm{\mu m}$ and $200\,\mathrm{\mu m}$.
The first peak corresponds to that of stellar emission, and the latter
corresponds to emission from ISM dust with $T\approx20\,\mathrm{K}$.
The excess emission of Model II between $\lambda=$~3--30~$\mathrm{\mu m}$
is due to warm dust in the accretion discs. The excess emission in
the far infrared ($\lambda>500\,\mathrm{\mu m}$) seen in Model II
is caused by the colder ($T<20\,\mathrm{K}$) dust. 

Assuming the distance to the cluster to be 140~pc, we have constructed
simulated images (Figure~\ref{fig:mips-images-limit}) for a \emph{SST}
MIPS (24, 70, and 160~$\mathrm{\mu m}$) observation using the
appropriate diffraction limits, the sensitivity limits and the angular
sizes of pixels. The emission at 24~$\mathrm{\mu m}$ traces the
locations of the stars and brown dwarfs very well. No clear
distinction can be made between the 70~$\mathrm{\mu m}$ image from
Model I and Model II. The major structures in the 160~$\mathrm{\mu m}$
images appear to be same in both Models, but the emission from the
isolated small structure in the lower half of the images is much
weaker in Model II. In the simulated 24~$\mathrm{\mu m}$ images, 8 out
of 50 objects are detected above the flux limit for Model I, and 18
out of 50 objects are detected for Model II. We note that 15 of the 27
brown dwarfs in the cluster (Model II) have $K<20$ and would therefore
detectable via deep imaging, while those that are too faint tend to be
the youngest, most deeply embedded sources.

Using the flux levels between 2.2~$\mathrm{\mu m}$ and 10~$\mathrm{\mu
  m}$, the spectral indices of each SEDs were computed. The objects
were then classified according to the spectral index values and the
ratio of $\log\lambda F_{\lambda}$ values at $\lambda=160\,\mathrm{\mu
  m}$ (\emph{SST} MIPS) and at $\lambda=2.2\,\mathrm{\mu m}$ ($K$ band).
We have found that 54 per cent of the objects are classified as
Class~0, none as Class~I, none as flat spectrum, 22 per cent as
Class~II and 24 per cent as Class~III for Model I. For Model II, 56
per cent of objects are classified as Class~0, 18 per cent as Class~I,
8 per cent as flat spectrum, 12 per cent as Class~II, and 6 per cent
as Class~III. We also found that the spectral index distribution of
Model~II ( with $K<20$ objects) is very similar to that of the
$\rho$~Ophiuchus cloud observation by \citet{greene:1994}.  Even
though our objects are $0.07\,\mathrm{Myr}$ old, there are a mixture
of Class 0--III objects; hence, the class does not necessary relate to
the ages of YSOs.

According to the simulated \emph{HKL} and mid-infrared (\emph{SST}
IRAC) colour-colour diagrams (Figures~\ref{fig:color-color-ukirt} and
\ref{fig:color-color-irac}), the disc objects (Model II) tend to be
slightly redder than discless objects (Model I) in the \emph{H-K} vs
\emph{K-L} diagram, but the two populations are not well separated.
This can be understood from the model cluster SEDs
(Figure~\ref{fig:sed_compare}) that show no significant difference in
the flux levels between $\lambda=$~1--3~$\mathrm{\mu m}$.  On the
other hand, the disc objects and discless objects are clearly
separated in the colour-colour diagram of the \emph{SST} IRAC bands
(Figure~\ref{fig:color-color-irac}). As expected, we find that longer
wavelengths are more efficient in detecting circumstellar discs. For
example mid-infrared ($\lambda=$~3--10~$\mathrm{\mu m}$) colours (e.g.
\emph{SST}~IRAC bands) are superior to \emph{HKL} colours. We find
that the intrinsic colours of the disc objects
(Figure~\ref{fig:color-color-irac}) show a distribution very similar
to that found by \citet{meyer:1997}.

The work presented in this paper can be extended to study how the
observable quantities (e.g. colours of stars in a young cluster)
evolves with time, by performing radiative transfer model calculations
with density structures from a hydrodynamics calculation at different
times (ages). Growth of dust grain sizes in accretion discs may become
important in predicting how the colours of objects evolve (e.g. see
\citealt{hansen:1974}; \citealt{wood:2002b}).

One of the obvious next steps in modelling star formation is to
combine SPH and radiative-transfer in order to more accurately predict
the temperature of the dust. It is interesting to note that the
computational expense of the radiative equilibrium calculation
performed here on one SPH time-slice (2400 CPU hours on UKAFF) is a
significant fraction of that required to perform the complete SPH
simulation (95\,000 CPU hours on the same machine) which involves may
thousands of time steps. Although it is not immediately clear how
often the temperature would have to be computed during a hydrodynamic
simulation (the radiative-transfer calculation would have its own
Courant condition), it is apparent that a much more efficient method
for calculating the radiation transfer is required before a combined
SPH/radiative-transfer simulation becomes tractable.

\section*{Acknowledgements}

We thank Mark McCaughrean for helpful discussions and a critical
reading of the manuscript. We are also grateful to the referee, Kenny
Wood. RK is supported by PPARC standard grant PPA/G/S/2001/00081. The
computations reported here were performed using the UK Astrophysical
Fluids Facility (UKAFF).

%
%----------------------------------------------------
% Bibliography follows here.
%-----------------------------------------------------
%\bibliographystyle{mn}
%\bibliography{apj-jour}

\label{lastpage}

%----------------------------------------------------
% Table 1 (Landscape figure) follows here
%----------------------------------------------------

\pagestyle{empty}
\setcounter{table}{0}

\begin{landscape}
\textwidth=240mm
\begin{table*}

\caption{Source catalogue}

\label{tab:catalog}

\scriptsize

\begin{tabular}{cccccccccccccccccc}
\hline 
Obj.$^{*}$&
$M^{*}$&
$L^{**}$&
$T_{\mathrm{{eff}}}^{**}$&
$R^{\dagger}$&
$A_{V}$&
$K^{\ddagger}$&
$F_{8.0}^{\ddagger}$&
$F_{24}^{\ddagger}$&
$a^{\ddagger}$&
$\mathrm{Class^{\ddagger}}$&
$A_{V}$&
$K^{\ddagger}$&
$F_{8.0}^{\ddagger}$&
$F_{24}^{\ddagger}$&
$a^{\ddagger}$&
$\mathrm{Class^{\ddagger}}$&
$i$\tabularnewline
&
$\left[\mathrm{M_{\sun}}\right]$&
$\left[\mathrm{L_{\sun}}\right]$&
$\left[K\right]$&
$\left[\mathrm{R_{\sun}}\right]$&
&
&
$\left[\mathrm{Jy}\right]$&
$\left[\mathrm{Jy}\right]$&
&
&
&
&
$\left[\mathrm{Jy}\right]$&
$\left[\mathrm{Jy}\right]$&
&
&
$\left[\mathrm{deg.}\right]$\tabularnewline
&
&
&
&
&
Model I&
\ldots{}&
\ldots{}&
\ldots{}&
\ldots{}&
\ldots{}&
Model II&
\ldots{}&
\ldots{}&
\ldots{}&
\ldots{}&
\ldots{}&
\ldots{}\tabularnewline
\hline 
1&
0.303&
1.002&
3740&
 2.4&
          3&
 8&
$5.8\times10^{-2}$&
$8.2\times10^{-3}$&
 $-2.4$&
 III&
 20000&
 11&
$1.4\times10^{-2}$&
$6.3\times10^{-1}$&
    $-0.7$&
    II&
86\tabularnewline
2&
0.211&
0.451&
3380&
 2.0&
           100&
 10&
$1.4\times10^{-2}$&
$5.5\times10^{0}$ &
    $-1.8$&
   III&
14000&
 13&
$7.5\times10^{-3}$ &
$1.9\times10^{0}$ &
 $0.1$&
 flat&
 87\tabularnewline
3&
0.731&
4.134 &
4610&
 3.2&
 20000&
 17&
$1.2\times10^{-2}$&
$3.5\times10^{-2}$ &
     $0.7$&
     0&
 20000&
 17&
$2.5\times10^{-4}$ &
$6.4\times10^{-2}$ &
 0.1&
     0&
 13\tabularnewline
4&
0.162&
0.273&
3200&
 1.7&
67&
 16&
$2.1\times10^{-3}$ &
$7.7\times10^{-4}$&
    $-0.7$&
    II&
 67&
 17&
$1.8\times10^{-2}$ &
$3.1\times10^{-1}$ &
    $1.8$&
     I&
53\tabularnewline
5&
0.012&
0.004&
2510&
 0.31&
 180&
 21&
$3.1\times10^{-5}$ &
$4.8\times10^{-5}$ &
    $-0.7$&
     0&
 970&
 21&
$8.4\times10^{-5}$ &
$2.0\times10^{-2}$ &
 $0.6$&
     0&
86\tabularnewline
6&
0.015&
0.005&
2540&
 0.36&
 66&
 20&
$7.6\times10^{-5}$ &
$9.1\times10^{-2}$ &
    $-0.2$&
     0&
 2500&
 19&
$4.7\times10^{-3}$ &
$2.9\times10^{-1}$ &
     $2.8$&
     0&
88\tabularnewline
7&
0.536 &
2.702&
4360&
 2.9&
             4&
 9&
$8.7\times10^{-2}$ &
$2.3\times10^{-1}$ &
   $-2.2$&
 III&
     4&
 9&
$1.2\times10^{0}$ &
$9.1\times10^{-1}$ &
     $0.4$&
     I&
 38\tabularnewline
8&
0.236&
0.560&
3460&
 2.1&
4&
 9&
$8.7\times10^{-2}$ &
$2.3\times10^{-1}$ &
 $-2.2$&
 III&
4&
 10&
$1.1\times10^{0}$ &
$1.0\times10^{0}$ &
  $1.1$&
     I&
 22\tabularnewline
9&
0.159&
0.264&
3180&
 1.7&
             7&
 10&
$2.0\times10^{-2}$ &
$2.9\times10^{-3}$&
 $-2.2$&
   III&
7&
 10&
$4.7\times10^{-1}$ &
$2.3\times10^{0}$ &
     $0.6$&
     I&
63\tabularnewline
10&
0.413&
1.888&
4090&
 2.7&
         20000&
 9&
$7.6\times10^{-2}$ &
$3.3\times10^{-2}$ &
    $-2.1$&
   III&
         20000&
 9&
$7.3\times10^{-2}$ &
$6.7\times10^{-2}$ &
    $-2.1$&
   III&
 16\tabularnewline
11&
0.260&
0.690&
3550&
 2.2&
         11000&
 21&
$4.5\times10^{-7}$ &
$2.0\times10^{-3}$ &
    $-4.1$&
     0&
 11000&
 22&
$2.0\times10^{-7}$ &
$1.2\times10^{-3}$ &
    $-3.6$&
     0&
 48\tabularnewline
12&
0.202&
0.417&
3360&
 1.9&
2&
 10&
$1.7\times10^{-2}$ &
$2.3\times10^{-3}$ &
    $-2.3$&
   III&
2&
 10&
$1.1\times10^{-1}$ &
$7.0\times10^{-1}$ &
    $-0.4$&
    II&
 52\tabularnewline
13&
0.025&
0.011&
2590&
 0.52&
           160&
 14&
$4.5\times10^{-4}$ &
$7.9\times10^{-5}$ &
    $-2.3$&
   III&
160&
 14&
$1.8\times10^{-3}$ &
$1.4\times10^{-2}$ &
    $-1.6$&
   III&
 25\tabularnewline
14&
0.032&
0.018&
2610&
 0.65&
             2&
 13&
$1.4\times10^{-3}$&
$2.2\times10^{-4}$ &
    $-2.2$&
   III&
             2&
 12&
$3.2\times10^{-2}$ &
$4.3\times10^{-2}$ &
    $-0.1$&
  flat&
 62\tabularnewline
15&
0.007&
0.002&
2470&
 0.24&
            19&
 15&
$2.6\times10^{-4}$&
$1.0\times10^{-4}$ &
    $-2.0$&
   III&
           460&
 15&
$1.6\times10^{-3}$ &
$3.6\times10^{-3}$ &
    $-0.0$&
  flat&
 86\tabularnewline
16&
0.012&
0.003&
2510&
 0.30&
18&
 16&
$3.5\times10^{-4}$ &
$6.7\times10^{-5}$ &
    $-1.4$&
    II&
18&
 15&
$8.2\times10^{-3}$ &
$8.2\times10^{-3}$ &
     1.0&
     I&
 43\tabularnewline
17&
0.008&
0.002&
2480&
 0.25&
19&
 16&
$2.3\times10^{-4}$ &
$4.5\times10^{-5}$ &
    $-1.4$&
    II&
19&
 16&
$5.4\times10^{-3}$ &
$7.1\times10^{-3}$ &
     $0.7$&
     I&
 72\tabularnewline
18&
0.008&
0.002&
2480&
 0.26&
6&
 15&
$3.0\times10^{-4}$ &
$2.9\times10^{-3}$ &
   $-1.8$&
   III&
             6&
 11&
$1.8\times10^{-1}$ &
$1.4\times10^{-1}$ &
     $0.6$&
     I&
 34\tabularnewline
19&
0.121&
0.155&
2930&
 1.5&
            30&
 11&
$1.6\times10^{-2}$ &
$5.6\times10^{-1}$ &
    $-1.9$&
   III&
30&
 11&
$9.9\times10^{-2}$ &
$1.7\times10^{0}$ &
    $-0.2$&
  flat&
 53\tabularnewline
20&
0.348&
1.381&
3910&
 2.6&
          4600&
 12&
$3.6\times10^{-2}$ &
$1.1\times10^{-2}$ &
  $-1.1$&
    II&
         4600&
 12&
$3.4\times10^{-2}$ &
$8.5\times10^{-3}$ &
 $-1.0$&
 II&
68\tabularnewline
21&
0.069&
0.066&
2670&
 1.2&
30&
 15&
$2.2\times10^{-3}$ &
$1.6\times10^{-1}$ &
    $-1.0$&
    II&
            31&
 15&
$5.3\times10^{-2}$ &
$3.4\times10^{-1}$ &
     $1.4$&
     I&
 71\tabularnewline
22&
0.114&
0.140&
2880&
 1.5&
          4600&
 14&
$4.7\times10^{-3}$ &
$3.9\times10^{-3}$ &
    $-1.0$&
    II&
4600&
 14&
$5.0\times10^{-3}$ &
$1.8\times10^{-3}$ &
    $-1.0$&
    II&
 70\tabularnewline
23&
0.032 &
0.018&
2610&
 0.67&
1000&
 44&
$5.0\times10^{-10}$ &
$7.0\times10^{-3}$ &
     $8.7$&
     0&
         1100&
 45&
$7.7\times10^{-8}$ &
$1.5\times10^{-3}$ &
     $7.5$&
     0&
79\tabularnewline
24&
0.173&
0.310&
3250&
 1.8&
           990&
 33&
$1.6\times10^{-8}$ &
$1.5\times10^{-2}$ &
     $4.1$&
 0&
           990&
 43&
$1.1\times10^{-9}$ &
$2.6\times10^{-3}$ &
     $5.5$&
     0&
78\tabularnewline
25&
0.229&
0.525&
3440&
 2.0&
          5200&
 33&
$2.4\times10^{-7}$&
$2.4\times10^{-3}$ &
     $4.0$&
     0&
          5200&
 45&
$1.1\times10^{-9}$ &
$1.1\times10^{-3}$ &
     $5.5$&
     0&
 52\tabularnewline
26&
0.133&
0.185&
3010&
 1.6&
          4700&
 35&
$6.3\times10^{-8}$ &
$9.0\times10^{-2}$ &
     $7.1$&
     0&
          8000&
 46&
$3.8\times10^{-11}$ &
$7.8\times10^{-4}$ &
     $3.5$&
     0&
 85\tabularnewline
27&
0.005&
0.002&
2460&
 0.22&
           220&
 27&
$6.3\times10^{-7}$ &
$3.6\times10^{-4}$ &
     $0.0$&
     0&
           220&
 30&
$6.8\times10^{-8}$ &
$1.6\times10^{-5}$ &
     $0.7$&
     0&
30\tabularnewline
28&
0.017&
0.006&
2550&
 0.38&
          1300&
 21&
$2.4\times10^{-6}$ &
$2.1\times10^{-5}$ &
    $-2.7$&
     0&
1300&
 19&
$3.0\times10^{-5}$ &
$3.1\times10^{-6}$ &
    $-1.9$&
   III&
 80\tabularnewline
29&
0.058&
0.050&
2650&
 1.1&
           340&
 18&
$4.1\times10^{-6}$ &
$2.9\times10^{-2}$ &
    $-3.9$&
     0&
           340&
 19&
$5.1\times10^{-6}$ &
$1.8\times10^{-2}$ &
    $-3.0$&
     0&
61\tabularnewline
30&
0.069&
0.065&
2670&
 1.2&
37&
 15&
$1.9\times10^{-4}$ &
$2.9\times10^{-3}$ &
    $-0.8$&
    II&
            37&
 17&
$1.7\times10^{-2}$ &
$4.6\times10^{-1}$ &
     $2.7$&
     0&
58\tabularnewline
31&
0.023&
0.010&
2590&
 0.50&
           120&
 17&
$3.2\times10^{-4}$ &
$1.4\times10^{-1}$ &
    $-0.8$&
     0&
           120&
 17&
$5.6\times10^{-2}$ &
$3.7\times10^{-1}$ &
     $2.7$&
     0&
 74\tabularnewline
32&
0.239&
0.571&
3470&
 2.1&
         22000&
 21&
$2.2\times10^{-3}$ &
$4.2\times10^{-2}$ &
     $1.9$&
     0&
         22000&
 21&
$2.3\times10^{-3}$ &
$2.6\times10^{-2}$ &
     $1.8$&
     0&
 65\tabularnewline
33&
0.087&
0.093&
2720&
 1.4&
           310&
 32&
$2.6\times10^{-6}$ &
$5.3\times10^{-3}$ &
     $3.5$&
     0&
           310&
 34&
$3.1\times10^{-6}$ &
$3.2\times10^{-3}$ &
     $5.7$&
     0&
 77\tabularnewline
34&
0.047&
0.035&
2630&
 0.90&
            21&
 14&
$1.6\times10^{-3}$ &
$2.5\times10^{-2}$ &
    $-1.4$&
    II&
7400&
 17&
$1.9\times10^{-4}$ &
$6.2\times10^{-2}$ &
     $0.4$&
     0&
 89\tabularnewline
35&
0.083&
0.086&
2710&
 1.3&
        51000&
 13&
$3.8\times10^{-3}$ &
$2.3\times10^{-2}$ &
    $-1.4$&
    II&
 51000&
 16&
$6.3\times10^{-4}$ &
$7.7\times10^{-2}$ &
    $-0.3$&
    II&
 66\tabularnewline
36&
0.044&
0.032&
2630&
 0.86&
15&
 14&
$1.7\times10^{-3}$ &
$6.0\times10^{-1}$ &
    $-1.4$&
    II&
15&
 14&
$4.4\times10^{-2}$ &
$2.6\times10^{-1}$ &
     $1.0$&
     I&
 75\tabularnewline
37&
0.022&
0.009&
2580&
 0.47&
           820&
 15&
$4.4\times10^{-4}$ &
$1.9\times10^{-3}$&
    $-1.6$&
    II&
           820&
 15&
$4.4\times10^{-4}$ &
$5.5\times10^{-4}$ &
    $-1.6$&
    II&
44\tabularnewline
38&
0.079&
0.081&
2690&
 1.3&
          3300&
 41&
$2.3\times10^{-9}$ &
$3.9\times10^{-3}$ &
     $5.0$&
     0&
          3300&
 53&
$1.3\times10^{-6}$ &
$2.3\times10^{-3}$ &
    $15.7$&
     0&
 61\tabularnewline
39&
0.070&
0.066&
2670&
 1.2&
450&
 50&
$1.8\times10^{-7}$ &
$8.4\times10^{-2}$ &
    $15.4$&
     0&
           450&
 64&
$5.0\times10^{-11}$ &
$2.9\times10^{-4}$ &
    $14.1$&
     0&
78\tabularnewline
40&
0.039&
0.025&
2620&
 0.77&
          1000&
 51&
$1.1\times10^{-7}$ &
$5.8\times10^{-2}$ &
    $16.0$&
     0&
          1000&
 61&
$5.9\times10^{-11}$ &
$8.9\times10^{-6}$ &
    $12.0$&
     0&
60\tabularnewline
41&
0.047&
0.035&
2630&
 0.90&
          1400&
 79&
$6.3\times10^{-8}$ &
$8.1\times10^{-2}$ &
    $32.1$&
     0&
1400&
 64&
$4.0\times10^{-11}$ &
$1.2\times10^{-3}$ &
    $14.4$&
     0&
63\tabularnewline
42&
0.095&
0.106&
2760&
 1.4&
          2300&
 50&
$2.3\times10^{-7}$ &
$3.7\times10^{-2}$ &
    $15.1$&
     0&
          2300&
 35&
$1.2\times10^{-7}$ &
$2.3\times10^{-2}$ &
     $5.8$&
     0&
 76\tabularnewline
43&
0.022&
0.009&
2580&
 0.48&
           580&
 49&
$1.4\times10^{-9}$ &
$4.1\times10^{-3}$ &
     $9.7$&
     0&
           580&
 40&
$1.2\times10^{-6}$ &
$1.9\times10^{-3}$ &
     $8.5$&
     0&
 77\tabularnewline
44&
0.102&
0.118&
2800&
 1.5&
          2700&
 53&
$2.5\times10^{-6}$ &
$2.4\times10^{-2}$ &
    $17.4$&
     0&
          2700&
 37&
$4.1\times10^{-7}$ &
$1.5\times10^{-2}$ &
     $6.2$&
     0&
 68\tabularnewline
45&
0.083&
0.087&
2710&
 1.3&
          3800&
 69&
$6.0\times10^{-13}$ &
$4.2\times10^{-3}$ &
    $21.0$&
     0&
          3800&
 36&
$1.4\times10^{-6}$ &
$2.3\times10^{-3}$ &
     $6.9$&
     0&
66\tabularnewline
46&
0.031&
0.017&
2610&
 0.64&
         1100&
 39&
$2.4\times10^{-9}$ &
$1.1\times10^{-3}$ &
 $5.1$&
     0&
          6100&
 44&
$8.4\times10^{-11}$ &
$1.6\times10^{-8}$ &
 $0.6$&
 0&
89\tabularnewline
47&
0.035&
0.021&
2620&
 0.70&
          1800&
 34&
$2.0\times10^{-9}$ &
$4.1\times10^{-4}$ &
     $0.2$&
     0&
          6100&
 35&
$3.1\times10^{-8}$ &
$5.0\times10^{-4}$ &
     $2.5$&
     0&
87\tabularnewline
48&
0.029&
0.015&
2610&
 0.60&
          1100&
 50&
$1.9\times10^{-10}$ &
$1.4\times10^{-7}$ &
     $6.2$&
     0&
          1100&
 40&
$1.8\times10^{-10}$ &
$2.6\times10^{-8}$ &
     1.5&
     0&
 56\tabularnewline
49&
0.013&
0.004&
2520&
 0.32&
         18000&
 65&
$1.4\times10^{-12}$ &
$3.5\times10^{-9}$ &
    $12.9$&
     0&
         19000&
 99&
$2.3\times10^{-16}$ &
$2.8\times10^{-8}$ &
    $26.5$&
     0&
 88\tabularnewline
50&
0.008&
0.002&
2480&
 0.25&
          7400&
 46&
$1.2\times10^{-8}$ &
$2.7\times10^{-2}$ &
    $12.9$&
     0&
         7400&
 41&
$8.2\times10^{-9}$ &
$3.4\times10^{-2}$ &
 $8.5$&
 0&
 63\tabularnewline
\hline
\end{tabular}

\vspace{0.3cm}

\raggedright

$\left(*\right)$~From the SPH calculation of \citet{bate:2003a}. 

$\left(**\right)$~Computed (at an age of 1/4~Myrs) from the isochrone data available on
\url{http://www.mporzio.astro.it/~dantona/prems.html}. See also \citet{d-antona:1998}. 

$\left(\dagger\right)$~Estimated by using the $L$ (column 3) and
$T_{\mathrm{eff}}$ (column 4) in the Stefan-Boltzmann law.

$\left(\ddagger\right)$~Based on the SEDs of individual objects
computed for an observer on $+z$ axis at 140\,pc. See section~\ref{sub:sed-individual}.

\normalsize

\end{table*}

\end{landscape}

%-------------------------------------------------------------
% The end of Table 1
%--------------------------------------------------------------

\end{document}